\documentclass{statsoc}

\usepackage{amsbsy}
\usepackage{amsmath}
\usepackage{amssymb}
\usepackage{multirow}
\usepackage{graphicx}
\usepackage{varwidth}
\usepackage{rotating}
\usepackage{hyperref}
\usepackage{color,xcolor, colortbl}
\usepackage{mathtools}
\usepackage{bm}
\usepackage[nolists, tablesfirst]{endfloat}
\newcommand{\bS}{\textbf{S}}
\newcommand{\bW}{\textbf{W}}
\newtheorem{theorem}{Theorem}
\newtheorem{lemma}{Lemma} 
\newtheorem{proposition}{Proposition}

\newtheorem{remark}{Remark}

\def\blackbox{\hfill\rule{.05in}{.05in}}
\newcommand{\bX}{\boldsymbol{X}}
\newcommand{\by}{\boldsymbol{y}}
\newcommand{\bV}{\boldsymbol{V}}
\newcommand{\bY}{\boldsymbol{Y}}

\newcommand{\bJ}{\bm{J}}
\newcommand{\bH}{\bm{H}}
\newcommand{\bI}{\bm{I}}

\newcommand{\bv}{\bm{v}}
\newcommand{\bj}{\bm{j}}
\newcommand{\bx}{\boldsymbol{x}}
\newcommand{\bs}{\bm{s}}

\newcommand{\btheta}{\boldsymbol{\theta}}
\newcommand{\blambda}{\bm{\lambda}}
\newcommand{\bTheta}{\boldsymbol{\Theta}}

\newcommand{\brho}{\bm{\rho}}

\newcommand{\bpsi}{\boldsymbol{\psi}}
\newcommand{\bvarepsilon}{\boldsymbol{\varepsilon}}

\title[CIF]{Scalable and Efficient Statistical Inference with Estimating Functions in the MapReduce Paradigm for Big Data}

\author[ZHOU \& SONG]{Ling Zhou and Peter X.-K. Song}
\address{University of Michigan, Ann Arbor, U.S.A.}

\begin{document}

\begin{abstract}
The theory of statistical inference along with the strategy of divide-and-conquer for large- scale data analysis has recently attracted considerable interest due to great popularity of the MapReduce programming paradigm in the Apache Hadoop software framework. The central analytic task in the development of statistical inference in the MapReduce paradigm pertains to the method of combining results yielded from separately mapped data batches. One seminal solution based on the confidence distribution has recently been established in the setting of maximum likelihood estimation in the literature. This paper concerns a more general inferential methodology based on estimating functions, termed as the Rao-type confidence distribution, of which the maximum likelihood is a special case. This generalization provides a unified framework of statistical inference that allows regression analyses of massive data sets of important types in a parallel and scalable fashion via a distributed file system, including longitudinal data analysis, survival data analysis, and quantile regression, which cannot be handled using the maximum likelihood method. This paper investigates four important properties of the proposed method: computational scalability, statistical optimality, methodological generality, and operational robustness. In particular, the proposed method is shown to be closely connected to Hansen's generalized method of moments (GMM) and Crowder's optimality. An interesting theoretical finding is that the asymptotic efficiency of the proposed Rao-type confidence distribution estimator is always greater or equal to the estimator obtained by processing the full data once. All these properties of the proposed method are illustrated via numerical examples in both simulation studies and real-world data analyses.
\end{abstract}

\section{Introduction}

\begin{figure}[h]
	\fbox{
		\begin{varwidth}{\dimexpr\textwidth-2\fboxsep+2\fboxrule\relax}	
			\hspace{0cm}\includegraphics[width = 17cm, height = 10cm]{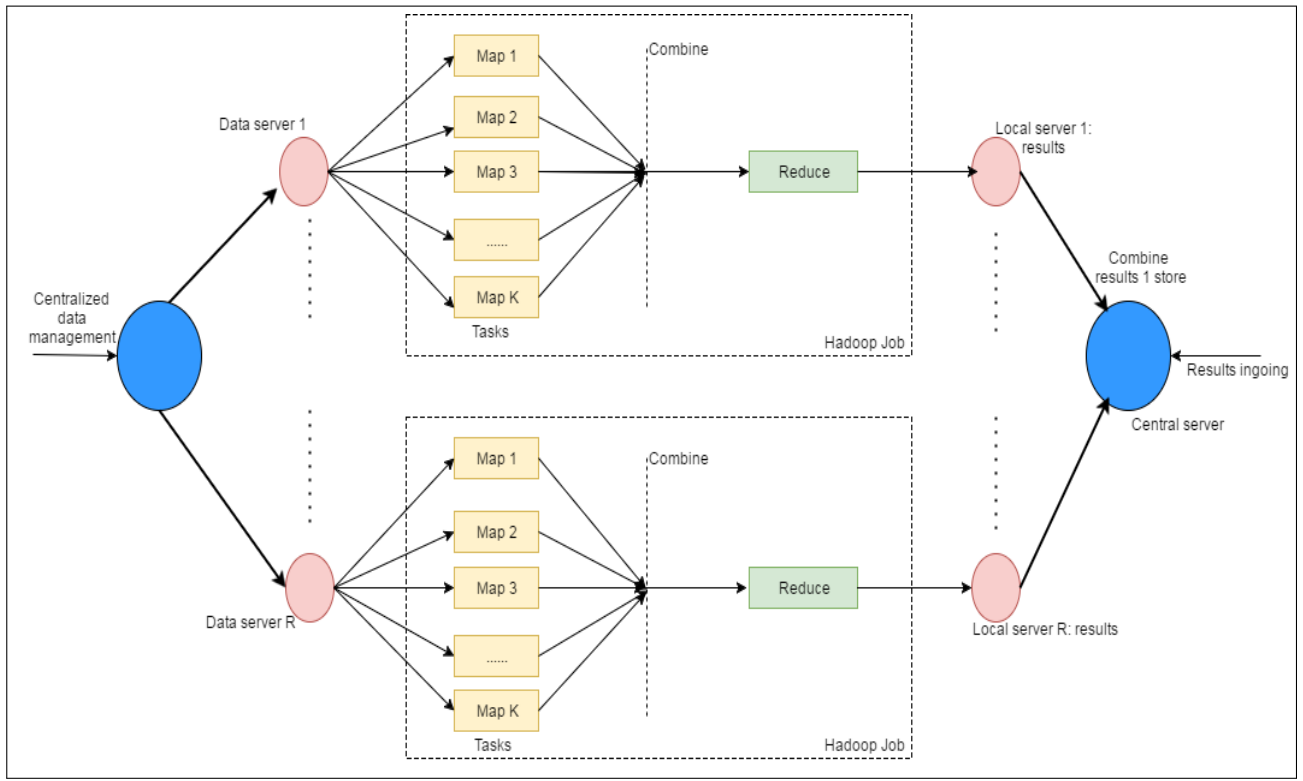}
	\end{varwidth}}
	\caption{Scematic flow chart of multi-server distributed data management and processing according to MapReduce paradigm as the heart of Hadoop platform.}
	\label{fig:me1}
\end{figure}

In response to rapidly growing demands of big data analytics and computational tools, parallel computing and distributed data storage have become the leading innovations for solving big data problems. For instance, multicore and cloud computing platforms, including the popular open source  \cite{hadoop}, are now the standard software technology used extensively in academia and industry \cite[]{user}.  This new distributed file system necessitates developing general statistical methodology that allows for analysing massive data through parallel and scalable operations in the Hadoop software framework. Being the core of Hadoop programming, MapReduce \cite[]{dean2008mapreduce, lammel2008google} represents a computing architecture that provides fast processing of massive  data sets.  Built upon the strategy of divide-and-conquer, the MapReduce paradigm refactors data processing into two primitives:  a {\em map} function, written by the user, to process distributed local data batches and generate intermediate results, and  a {\em reduce} function, also written by the user, to combine all intermediate results and then generate summary outputs.  The detailed examples and implementation are referred to \cite{dean2008mapreduce}.  Figure~\ref{fig:me1} displays a schematic outline of MapReduce workflow, which splits data and performs computing tasks in the form of parallel computation.  The salient features of MapReduce include scalability and independence of data storage; the former enables automatic parallelization and allocation of large-scale computations, and the latter allows to process data without requiring it to be loaded into a common data server.  In this way, it avoids the high computational cost of loading input data into a centralized data server prior to the analysis.

Although MapReduce and its variants have shown superb power to process large-scale data-intensive applications on high-performance clusters, most of these systems are restricted with an acyclic data flow,  which is not suitable for general statistical analysis in that iterative numerical operations are involved \cite[]{yang2007map}. This is because 
operation of an iterative algorithm, like Newton-Raphson, requires repeatedly reload data from multiple data disks into a common server, incurring a significant performance penalty \cite[]{zaharia2010spark}.
This paper is motivated to address this computational hurdle by a new combined inference approach in the framework of confidence distributions, so the resulting methodology of statistical estimation and inference can avoid repeated operations of data reloading in iterative numerical jobs, and truly enjoys the power of scalability offered by a distributed file system such as Hadoop. 

In most of Hadoop platforms, data partition and allocation in the Map-step may be operated by certain built-in system software to accommodate specific hardware configurations. As far as statistical inference concerns, methodological needs take place mostly in the Reduce-step, in which statistical approaches of combining separate results are called for.  Unlike the ordinary least squares method in the linear model, most of nonlinear regression models are relied on certain iterative numerical algorithms to obtain point estimates and quantities for statistical inference. Technically, these iterative algorithms typically request processing the entire data under centralized non-separable calculations, except for some simple linear operations of data, such as arithmetic mean, count and proportion. Repeatedly loading local datasets into a central data server is not only extremely time-consuming, but is prohibited if data volume exceeds the memory, or if, physically, data sets are stored in different servers located in different sites with no data merging agreement in place. This presents indeed the challenge of generalizing the MapReduce paradigm to many important statistical models and their numerical toolboxes, such as generalized estimating equations for longitudinal data, Cox proportional hazards model for survival data, and quantile regression, among others.    

We consider a class of parametric models $\mathcal{P}_{\btheta} = \{p_{\btheta}(w), \btheta \in \Theta\}$, with the parametric space $\Theta \subseteq R^p$. Here $p$ is assumed to be fixed. In many important statistical problems where the underlying probability density function $p_{\btheta}(w)$ by which the data are generated cannot by fully specified, estimating functions built upon some aspects of the probability mechanism such as moment conditions are utilized to carry out parameter estimation and inference.  Given independent samples $\bW_{i}, i = 1, \dots, n $, in the current literature, parameter $\btheta$ may be estimated as a solution, denoted by $\hat{\btheta}$,  of the following estimating equation: 
\begin{eqnarray}
\label{met:eq}
\psi_{full}(\bW; \btheta) \overset{def}{=} n^{-1}\sum_{i=1}^{n}\psi(\bW_{i}; \btheta) = \bm{0};
\end{eqnarray}
that is, $\psi_{full}(\bW; \hat{\btheta}) = \bm{0}$, where $\bW = \left\{\bW_{1}, \cdots, \bW_{n}\right\}$ denotes the entire data.  See for example \cite{mcleish2012theory, heyde2008quasi}, and \citet[Chapter 3]{song2007correlated} for the theory of estimating functions, and more references therein.  When $\psi(\cdot)$ in (\ref{met:eq}) is a score function, namely the first-order derivative of a log-likelihood, the solution $\hat{\btheta}$ is the maximum likelihood estimator (MLE). In other words, the method of MLE may be regarded as a special case of the estimating function method.  In general, equation (\ref{met:eq}) encompasses many important cases, such as the generalized estimating equation (GEE) for longitudinal data, the partial likelihood score function in the Cox model for survival data, and the quantile regression estimating function, and so on.  If there were a ``God-made'' computer with infinite power, there would be no pressing need of developing new methods for processing big data, and the existing methodologies and associated software would be directly applied to solve equation (\ref{met:eq}).  Unfortunately, thus far human-made computer does not have such capacity, so the MapReduce paradigm that implements the divide-and-conquer strategy has emerged as one of state-of-the-art computational solutions to make big data computation feasible.  Using this computing platform to implement divide-and-conquer strategy for statistical estimation and inference leads to two primary methodological questions: 
\begin{itemize}
	\item[(a)] If is it possible,  and if so how, to establish a statistical inference procedure that is suitable to implement the Reduce-step in the MapReduce paradigm? Specially,  consider a data partition scheme to, \emph{say}, $K$ disjoint subsets, $\bW = \cup_{k=1}^K\bW^{(k)}$. In the Map-step each sub-dataset $\bW^{(k)}$ is processed in a parallelized fashion,  where equation (\ref{met:eq}) is solved separately at individual computer nodes by existing statistical software (\emph{e.g.} R package \verb|gee|), resulting in estimates $\hat{\btheta}_k$, $k=1,\ldots,K$. Then,  in the Reduce-step, there is a need of developing a procedure to gather these separate estimates $\hat{\btheta}_k$ and their variances to perform a valid statistical inference, if possible.  
	\item[(b)] Suppose that there exists an established procedure in part (a) that enables to derive a combined or meta estimator, {\em say}, $\hat{\btheta}_{meta}$.  Then, there is a need of assessing the performance of the proposed $\hat{\btheta}_{meta}$, in terms of its estimation bias, estimation efficiency, and robustness, and comparing it to the solution of equation (\ref{met:eq}) obtained by processing the entire data once using a ``God-made'' computer.  For convenience, the latter solution, denoted as $\hat{\btheta}_{full}$,  serves as the benchmark solution in the rest of this paper.  
\end{itemize}

Solutions to these two questions above have been discussed  in the setting of maximum likelihood estimation in the literature. Recently, \cite{lin2010relative} and \cite{liu2015multivariate} proposed meta estimators, $\hat{\btheta}_{meta}$, defined as an inverse variance weighted average of $\hat{\btheta}_k$, with  $\hat{\btheta}_k$ being MLE obtained from each sub-dataset $\bW^{(k)}$. \cite{liu2015multivariate} showed that their meta estimator is asymptotically as efficient as the MLE derived from using the entire dataset once. In the setting of random-effects models, \cite{zeng2015ran} reported a similar finding; that is,  their meta estimator is at least as efficient as the one obtained from the entire data. The divide-and-conquer scheme has been also studied in other statistical problems, such as hypothesis testing; see also 
\cite{battey2015distributed, chen2014split, lee2017communication, li2013statistical, zhang2015divide, zhao2016partially}, among others. 

One of the most promising approaches to statistical inference suitable for the Reduce-step is the so-called {\em confidence distribution} \cite[]{xie2013confidence}, which was the method that has been applied by \cite{liu2015multivariate} to derive an asymptotically fully efficient solution. 
As a ``distributional estimator", confidence distribution (CD) has gained increasing attention due to its computational convenience, giving rise to a theoretically unified framework for estimation and statistical inference. The concept of confidence distribution may be traced back to \cite{mr1763essay}, \cite{fisher1930inverse, fisher1956statistical} and later \cite{efron1993bayes} in the field of fiducial inference; see 
\cite{xie2013confidence} for a comprehensive review, and more references therein. Relevant to this paper, the most critical question here is why the CD approach suits for the derivation of a combined estimation and inference in the Reduce-step.  The key insight learned from the setting of maximum likelihood estimation lies in the fact that the construction of CD only requires summary statistics rather than individual subject-level data, and that the resulting inference has shown asymptotically no loss of statistical power. This is aligned with the analytic goal that the Reduce step aims to achieve. Thus, in this paper we consider generalizing the CD approach to the setting of estimating functions through which we hope to integrate separate pieces of inferential information to obtain a combined inference in mathematical rigor. This is different from the currently popular strategy of directly combining estimators. Our proposed Reduce-step can be applied to deal with a broad range of statistical models especially in cases where likelihood is not available. 

To facilitate our discussion, we begin with a simple heuristic presentation of the CD approach in the Reduce-step.  Suppose that under some regularity conditions such that the standard large-sample properties hold, for each sub-dataset, estimator $\hat{\btheta}_k$ satisfies $n^{1/2}_k(\hat{\btheta}_k-\btheta_{k, 0}) \stackrel{asy.}{\sim} \mathcal{N}(0, \bj^{-1}_k(\btheta_{k, 0})), k=1,\ldots,K,$ where $\btheta_{k, 0}$ is the underlying true parameter and $\bj_k$ is the Godambe information or $\bj_k^{-1}$ is the sandwich covariance matrix. Then $\Phi\left(n_k^{1/2}\bj_k^{1/2}(\btheta_{k, 0}) (\hat{\btheta}_{k} - \btheta_{k, 0})\right)$ follows asymptotically the $p$-dimensional independent copula or the $p$-dimensional distribution of independent uniform marginals, where $\Phi$ is the $p$-variate normal cumulative distribution function with mean $\bm{0}$ and the identity variance matrix.  According to \cite{fisher1935fiducial}, it constitutes a pivotal quantity, a distributional element essential for the so-called fiducial inference.  The pivotal density is termed as the confidence density by \cite{efron1993bayes} whose expression takes the form, $h_k(\btheta) \propto \exp\left\{-n_k (\hat{\btheta}_{k} - \btheta)^T \bJ_{n_k}(\hat{\btheta}_k) (\hat{\btheta}_{k} - \btheta)\right\}$, where $\bJ_{n_k}$ is a consistent estimate of the information matrix $\bj_k$. Clearly, the above confidence density $h_k(\btheta)$ may be used to construct confidence regions of $\btheta$ at any given confidence level. Suggested by \cite{singh2005combining}, a meta estimator of 
$\btheta_{0}$, under the homogeneity assumption $\btheta_{k, 0} \equiv \btheta_0, k = 1, \dots, K$,  may be  obtained by maximizing a joint confidence density of the following form: 
$$
\arg\max_{\btheta}\prod_{k=1}^K h_k(\btheta) =  \arg\max_{\btheta}\prod_{k=1}^K \exp\left\{-n_k (\hat{\btheta}_{k} - \btheta)^T \bJ_{n_k}(\hat{\btheta}_k) (\hat{\btheta}_{k} - \btheta)\right\}, 
$$
where the $K$-fold product is due to the independence across the $K$ sub-datasets. 
This procedure has been thoroughly discussed by \cite{liu2015multivariate} in the context of maximum likelihood method where matrix $\bJ_{n_k}$ is the observed Fisher information matrix, a special case of the Godambe information matrix $\bj_k$ when the Bartlett identity holds \cite[Chapter 3]{song2007correlated}. 

In the setting of estimating functions, there is another way to establish the asymptotic normality, based directly on the estimation functions, $n^{1/2}_k\psi_{k\_sub}(\btheta_0) \overset{def}{=} n^{-1/2}_k\sum_{i=1}^{n_k}\psi(\bW_{i}^{(k)}; \btheta_0) \stackrel{asy.}{\sim}  \mathcal{N}\left(\bm{0}, \bv_k(\btheta_{0})\right)$, where matrix $\bv_k$ is the variability matrix, {\em i.e.}, the variance of the estimating function $\psi$, different from the sandwich covariance matrix $\bj^{-1}_k$ above. Likewise, we may construct another pivotal quantity $\Phi\left( n^{1/2}_k\bV^{-1/2}_{n_k}(\hat{\btheta}_k)\psi_{k\_sub}(\btheta_0) \right)$ to obtain a different CD, where $\bV_{n_k}$ is a consistent estimate of the variability matrix $\bv_k$. \cite{godambe1978some} had strongly advocated the use of estimating functions $\psi_{k\_sub}(\cdot)$, instead of its estimator $\hat{\btheta}_k$, to make statistical inference due to better finite-sample performances.  This motivates us to take a new route of investigation to construct pivotal quantities and then confidence distributions, which results in a  different meta estimation.  In order to differentiate these two different routes of CD constructions, we borrow terms from the classical hypothesis testing theory, and name the estimator-driven CD as the {\em Wald-type CD} and our new estimating function based CD  as the {\em Rao-type CD}. Moreover, for the convenience of exposition, they are abbreviated as Wald-CD and Rao-CD, respectively, in this paper. There has been little work in the literature concerning MapReduce approaches to parameter estimation and inference with estimating functions;  \cite{lin2011aggregated} proposed an aggregated estimating equation (AEE), which was not developed in the CD framework, and thus it is less general in comparison to the proposed Rao-CD method; the detailed comparison between AEE and our Rao-CD is available in both methodology discussion and simulation studies later in this paper.    


The primary objective of this paper is to develop, assess and compare our proposed Rao-CD approach in the Reduce-step with existing methods. The focus of investigation will be on the following four aspects.  (i) Scalability. Being implemented by the strategy of divide-and-conquer within the MapReduce paradigm, the Rao-CD estimation and inference procedures are scalable to massive large-scale data.   (ii) Optimality. The Rao-CD approach is closely related to seminal Hansen's generalized method of moments (GMM), which supplies a powerful analytic tool for us to establish theoretical justifications for the optimality of the proposed method.  Moreover, the Rao-CD approach is also shown to be connected to the Crowder's optimality, another important perspective on the optimality of the Rao-CD meta estimation. The most interesting theoretical result in this paper is given by Theorem \ref{the:corie} in Section \ref{sec:com}; that is, the asymptotic estimation efficiency of the Rao-CD meta estimator is always equal or higher than that of the benchmark estimator $\hat{\btheta}_{full}$ obtained by processing the entire data once. (iii) Generality.  The proposed Rao-CD method provides a general valid inference procedure by combining results from separately fitted models where the likelihood is not available. It also includes \cite{lin2011aggregated}'s aggregated estimating equation (AEE) estimator as a special case. (iv) Robustness.  The proposed Rao-CD method is shown to be robust against certain heterogeneity and/or contaminated data, which has not been studied in any divide-and-conquer method from the frequentest perspective. The robustness is rooted in two facts: (a) diluted abnormality. Data partition in the Map-step may allocate abnormal data cases into some sub-datasets while the others contain no outliers. When this happens, the analysis would be only affected within a small number of sub-datasets, while analyses with the majority of sub-datasets remain unaffected. (b) Automatic down-weighting. The Rao-CD confidence density provides an automatic down-weighting scheme to minimize the contributions from bad estimators (with inflated variances) due to the fact that weighting is anti-proportional to the variance of an estimator \cite[]{qu2004assessing}. The consequent combined estimator in the Reduce-step will be robust by the two layers of protection. In contrast, if the entire data is run together in equation~\eqref{met:eq}, the analysis will be affected by a few strong influential data cases, unless certain robustness treatments are applied to estimating functions. 
In Bayesian inference, \cite{minsker2014robust} proposed a robust and scalable approach to Bayesian analysis in a big data framework.
	
The rest of the paper is organized as follows.  In Section~\ref{sec:cef}, we introduce the Rao-CD in detail. Section~\ref{sec:com} discusses the development of Rao-CD in the Reduce-step, including its optimality. Section~\ref{sec:the} shows the theoretical properties of CD meta estimators.  Section~\ref{sec:imp} focuses on a fast MapReduce implementation procedure. Section~\ref{sec:ex} presents three useful examples. The numerical performance is evaluated in Section~\ref{sec:simu}. 
Finally, we apply the Rao-CD method to several real-world data sets in Section~\ref{sec:real}, and we conclude with a discussion of the Rao-CD's limitations and future work in Section~\ref{sec:dis}. All the conditions and proofs are given in the Appendix.

\section{Rao-type confidence distribution}
\label{sec:cef}
When the likelihood is not available, the theory of estimating functions provides an appealing approach to obtain an estimator $\hat{\btheta}$ of $\btheta_0$, as an solution to equation~(\ref{met:eq}) \cite[\emph{e.g.}][]{heyde2008quasi}. In this theoretical framework, there exist two forms of information matrices, the variability matrix and sensitivity matrix, denoted by $\bv(\btheta) \overset{def}{=} \mbox{Var}\left\{n^{1/2}\psi_{full}(\bW; \btheta)\right\}$ and $\bs(\btheta) = -\partial E\left\{\psi_{full}(\bW; \btheta)\right\}/\partial \btheta$, respectively. Both of them are assumed to be positive definite in this paper. The Bartlett identity refers to the equality, $\bv(\btheta) =  \bs(\btheta)$ for $\btheta \in \bTheta$, which holds for the case of $\psi$ being the score function. Under some regularity conditions \cite[][Chapter 3]{song2007correlated}, the estimating function $\psi_{full}(\bW; \btheta_{0})$ has the following asymptotic normal distribution:
\begin{eqnarray}
\label{met:ad}
\sqrt{n}\psi_{full}(\bW; \btheta_{0}) \stackrel{d}{\rightarrow} \mathcal{N}\left(0, \bv(\btheta_{0})\right), \hspace{0.2cm}{as} \hspace{0.2cm} n\to \infty.
\end{eqnarray}
It follows that $\left\{\bV_{n}^{-1/2}(\btheta_0)\right\}\psi_{full}(\bW; \btheta_0) \stackrel{asy.}{\sim} \mathcal{N}\left(0, \bI\right)$, where $\bV_{n}(\btheta)$ is the sample variance, $\bV_{n}(\btheta) = n^{-1}\sum_{i=1}^{n}\psi(\bW_{i}; \btheta)\psi(\bW_{i}; \btheta)^{T}$, a root-$n$ consistent estimator of $\bv(\btheta)$. Denote $\hat{\bV}_{n} = \bV_{n}(\hat\btheta_{full})$. Under the same regularity conditions, it is also known that the estimator $\hat\btheta_{full}$ is asymptotically normally distributed,
\begin{eqnarray}
\label{met:adp}
\sqrt{n}\left(\hat{\btheta}_{full} - \btheta_{0}\right) \stackrel{d}{\rightarrow} \mathcal{N}\left(0, \bj^{-1}(\btheta_{0})\right), \hspace{0.2cm}{as} \hspace{0.2cm} n\to \infty,
\end{eqnarray}
where $\bj(\btheta) = \bs^{T}(\btheta)\bv^{-1}(\btheta)\bs(\btheta)$
is the Godambe information matrix.  

According to the definition of confidence distribution \cite[]{schweder2002confidence, singh2005combining} from the asymptotical normality in~(\ref{met:ad}), in this paper we define a Rao-type confidence distribution with respect to $\psi_{full}$ as follows:  
\begin{eqnarray}
\label{def:cee}
H_{R}(\btheta_{0}) \overset{def}{=} \Phi\left(n^{1/2}\hat{\bV}_{n}^{-1/2}\psi_{full}(\bW; \btheta_{0})\right),
\end{eqnarray}
where $\Phi$ is the $p$-variate normal cumulative distribution function with mean $\bm{0}$ and the identity variance matrix. Clearly, asymptotically, $H_{R}(\btheta_0) \sim u_1u_2\cdots u_p$, with $u_j \stackrel{iid}{\sim} \text{Unif}(0, 1), j = 1, \dots, p$. 
Likewise, from the asymptotic normality in~(\ref{met:adp}), a Wald-type confidence distribution for $\btheta_0$ is given by
\begin{eqnarray}
\label{def:cd}
H_{W}(\btheta_{0})  \overset{def}{=} \Phi\left(n^{1/2}\bJ^{1/2}_{n}(\hat{\btheta}_{full})(\hat{\btheta}_{full} - \btheta_{0})\right),
\end{eqnarray}
where $\bJ_{n}(\btheta) = \bS^{T}_{n}(\btheta)\bV^{-1}_{n}(\btheta)\bS_{n}(\btheta)$  is the observed Godambe information matrix with $\bS_{n}(\btheta)$ being the observed sensitivity matrix, namely, $\bS_{n}(\btheta) = -n^{-1}\sum_{i=1}^{n}\dot{\psi}(\bW_{i}; \btheta)$, provided that the estimating function $\psi(W; \btheta)$ is differentiable. Again, $H_{W}(\btheta_0) \sim u_1u_2\cdots u_p$, asymptotically. In the current literature, this type of Wald-CD \cite[]{liu2015multivariate, xie2013confidence} has been the choice of confidence distribution considered in the setting of maximum likelihood estimation.

A question arises naturally: what is the relationship between Rao-CD and Wald-CD? An answer may be drawn by an analysis resembling the classical comparison between Wald test and Score test in the theory of hypothesis testing; see for example \cite{engle1984wald}. From the definition of CD, both Wald-CD, $H_{W}(\btheta_0)$, and Rao-CD, $H_{R}(\btheta_0)$, are distributional estimators for statistical inferences, and they can be shown to be asymptotically equivalent for inference on $\btheta_{0}$ (see Theorem~\ref{the:cd&cef} and Lemma~\ref{lem:cd} in Appendix B.4). On the other hand, in comparison to Wald-CD, Rao-CD has the following advantages. First, Rao-CD is invariant under one-to-one parameter transformation, {\em say}, $\blambda = \blambda(\btheta)$, which leads to a different distribution function of parameter $\blambda$ but an equivalent estimating function \cite[]{godambe1991estimating}. 
Second, Rao-CD is favorable if calculation of the sensitivity matrix, $\hat{\bS}_n$, which is involved in Wald-CD, is analytically tedious or numerically unstable \cite[]{song2005maximization}. Note that Rao-CD only requires calculation of the variability matrix, $\hat{\bV}_n$. 
Third, as being the most important advantage, Rao-CD provides a much more convenient theoretical framework than Wald-CD to establish theoretical properties of the CD meta estimation, because we can shown that it is connected to Hansen's GMM in Section~\ref{sec:com}. Using this remarkable connection, we can provide theoretical justifications for optimal efficiency and estimation robustness against contaminated data of the CD meta estimator.  

\section{Rao-CD meta estimation from parallel datasets}
\label{sec:com}
\subsection{Definition}

In this section, we present the procedure of combining Rao-type confidence distributions to derive a meta estimator for a common parameter $\btheta$ of interest. Consider $K$ parallel datasets $\bW^{(1)}, \dots, \bW^{(K)}$, each with $n_k$ observations, $k = 1,\dots, K$. Assume these $K$ sub-datasets are independently sampled from $K$ disjoint sets of subjects, and each of which is processed separately by the estimating equation in~\eqref{met:eq}, leading to estimators $\hat{\btheta}_k$, $k = 1,\dots, K$. 

In a similar spirit to ideas given in \cite{singh2005combining} and \cite{liu2015multivariate}, the cross-dataset independence permits multiplication of the $K$ Rao-type confidence distributions given by~(\ref{def:cee}) with respect to  $\bW^{(k)}, k = 1, \dots, K$ . Specifically, we consider the density of Rao-type CD for the $k$-th sub-dataset $\psi_{k\_sub}$ by, subject to some asymptotically constant factors, $$h_{R, k}(\btheta; \hat{\bV}_{n_k}) \propto \phi\left\{n_k^{1/2}\hat{\bV}_{n_k}^{-1/2}\psi_{k\_sub}(\bW^{(k)}; \btheta)\right\},$$ where $\phi(\cdot)$ is a $p$-variate normal density function with mean $\bm{0}$ and the identity variance matrix, and $\hat{\bV}_{n_k} = \bV_{n_k}(\hat{\btheta}_k)$. To proceed the CD approach, we take a product of these $K$ confidence densities as follows,
\begin{eqnarray}
\label{met:cef}
h_{R}^c(\btheta) = \prod_{k=1}^{K}h_{R, k}(\btheta; \hat{\bV}_{n_k}).
\end{eqnarray}
Moreover, we define a meta estimator of $\btheta_0$ by $\hat{\btheta}_{rcd} = \arg\max_{\btheta}h_{R}^c(\btheta)$. We show in the next two subsections 3.2-3.3 that this meta estimator $\hat{\btheta}_{rcd}$ in \eqref{met:cef} has the following two important properties of optimality, namely the Crowder's optimality and the Hansen's optimality in the context of generalized method of moments estimator (GMM).

\subsection{Crowder's optimality}

Note that $\hat{\btheta}_{rcd} = \arg\max_{\btheta}\sum_{k=1}^{K}\log h_{R, k}(\btheta; \hat{\bV}_{n_k})$, which is obtained as a solution of the following estimating equation: 
\begin{eqnarray}
\label{eq:cef}
\Psi_{R}(\btheta) \stackrel{def}{=}n^{-1/2}\sum_{k=1}^{K}n_k\bS^{T}_{n_k}(\btheta)\hat{\bV}^{-1}_{n_k}\psi_{k\_sub}(\bW^{(k)}; \btheta) = \bm{0}.
\end{eqnarray}
Under some regularity conditions, \cite{crowder1987linear} showed that the optimal estimating function, $\Psi^*_c(\btheta)$, in the Crowder's class of estimating functions, $\mathcal{G}_c = \left\{\Psi_c(\btheta)\right\}$ of the following forms, $$\Psi_{c}(\btheta) = n^{-1/2}\sum_{k=1}^{K}n_kC_k(\btheta)\psi_{k\_sub}(\bW^{(k)}; \btheta),$$
is the one with $C^*_k(\btheta) = \bs^{T}_{k}(\btheta)\bv^{-1}_{k}(\btheta)$,  where $\bs_k$ and $\bv_k$ are the sensitivity and variability matrices of $\psi_{k\_sub}$ with sub-dataset $\bW^{(k)}$. Also see Theorem 3.13 in \cite{ song2007correlated}. The following proposition shows that the estimating function $\Psi_{R}(\btheta)$ in~(\ref{eq:cef}) is asymptotically equivalent to the Crowder's optimal estimating function $\Psi^*_c(\btheta) =n^{-1/2}\sum_{k=1}^{K}n_kC^*_k(\btheta)\psi_{k\_sub}(\bW^{(k)}; \btheta)$ at $\btheta = \btheta_{0}$. 
\begin{proposition}
\label{prop:crowder}
Under regularity conditions (C1)-(C3) and (C4.0) in Appendix A, if $K = O(n^{1/2 - \delta})$ with some $\delta \in (0, 1/2]$, we have the $l_2$-norm of $\Psi_{R}(\btheta_{0}) - \Psi^*_{c}(\btheta_0)$ asymptotically converges to 0; that is, $\|\Psi_{R}(\btheta_{0}) - \Psi^*_{c}(\btheta_0)\|_2 = o_p(1)$ as $\min_k{n_k} \rightarrow \infty$.
\end{proposition}

The proof of proposition~\ref{prop:crowder} is given in Appendix B.3. Proposition~\ref{prop:crowder} indicates that $\Psi_{R}(\btheta)$ in~\eqref{eq:cef} is asymptotically the optimal estimating function, in the sense that the resulting meta estimator $\hat{\btheta}_{rcd}$ has the largest Godambe information among those obtained with $\Psi_c(\btheta) \in \mathcal{G}_c$.

\subsection{Hansen's optimality}
Since $h_{R}^{c}(\btheta) \varpropto \exp\left\{-\sum_{k=1}^{K}\frac{n_k}{2}\psi_{k\_sub}(\bW^{(k)}; \btheta)^{T}\hat{\bV}_{n_k}^{-1}\psi_{k\_sub}(\bW^{(k)}; \btheta)\right\}$, it is interesting to see that the Rao-CD estimator $\hat{\btheta}_{rcd} = \arg\max_{\btheta}h^c_R(\btheta)$ is equivalent to minimizing the following quadratic function,
\begin{eqnarray}
\label{gmm:cef}
\hat{\btheta}_{rcd} = \arg\min_{\btheta}\left\{\bpsi_{n}(\bW; \btheta)^{T}\hat{\mathbb{V}}_{n}^{-1}\bpsi_{n}(\bW; \btheta)\right\},	
\end{eqnarray}
where $\bpsi_{n}(\bW; \btheta) = \left\{\sqrt{n_1}\psi_{1\_sub}(\bW^{(1)}; \btheta), \dots, \sqrt{n_k}\psi_{K\_sub}(\bW^{(K)}; \btheta)\right\}^{T}$ is an extended vector of estimating functions, each based on one sub-dataset, and $\hat{\mathbb{V}}_{n} = \mbox{block-diag}\left\{\hat{\bV}_{n_1}, \dots, \hat{\bV}_{n_K}\right\}$. Here $\bpsi_n(\bW; \btheta)$ is an over-identified estimating function in the sense that its dimension is bigger than the dimension of $\btheta$. Because of the independent sampling across the $K$ sub-datasets, the variance of $\bpsi_n(\bW; \btheta)$ will be block-diagonal, and $\hat{\mathbb{V}}_{n}$ is a consistent estimator of its variance. According to \cite{hansen1982large}, expression~(\ref{gmm:cef}) presents a form of GMM. Thus, it is known from the classical theory of GMM that under some regularity conditions, our proposed meta estimator $\hat{\btheta}_{rcd}$ has the smallest asymptotic variance among those meta estimators $\hat{\btheta}_{meta}$ given by the following forms:  
$$\hat{\btheta}_{meta} = \hat{\btheta}_{meta}(\mathbb{C}_n)  = \arg\min_{\btheta} \left\{\bpsi_{n}(\bW; \btheta)^{T}\mathbb{C}_n\bpsi_{n}(\bW; \btheta)\right\},$$ 
where $\mathbb{C}_n$ is a certain weighting matrix from, {\em say}, the class of semi-positive definite matrices. Expression \eqref{gmm:cef} provides a convenient theoretical framework for the development of large-sample properties for the proposed meta estimator $\hat{\btheta}_{rcd}$, as many established theorems and properties for the GMM may be applied here.

On the other hand, based on the asymptotic normality of estimators $\hat{\btheta}_k, k = 1, \dots, K$, the Wald-CD meta estimator takes the following form,
\begin{eqnarray}
\label{gmm:cd}
\hat{\btheta}_{wcd} = \arg\min_{\btheta} \left\{\sum_{k=1}^{K}n_k(\hat{\btheta}_k - \btheta)^{T}\hat{\bS}^{T}_{n_k}\hat{\bV}^{-1}_{n_k}\hat{\bS}_{n_k}(\hat{\btheta}_k - \btheta)\right\},
\end{eqnarray}
which is obtained using a product of the Wald-type confidence distributions in~(\ref{def:cd}), where $\hat{\bS}_{n_k} \stackrel{def}{=}\bS_{n_k}(\hat{\btheta}_k)$ is the estimated sensitivity matrix. Applying similar arguments established in the classical theory of hypothesis testing for a comparison between Rao's score test and Wald's test,  we can show that $\hat{\btheta}_{rcd}$ in~(\ref{gmm:cef}) and $\hat{\btheta}_{wcd}$ in~(\ref{gmm:cd}) are indeed asymptotically equivalent under some smooth conditions of estimating function $\psi$ such as condition (C4.2) in Appendix A. See Theorem~\ref{the:cd&cef} for the details.

\section{Large sample properties}
\label{sec:the}

\subsection{Consistency and asymptotic normality}

We establish the consistency and asymptotic normality of $\hat{\btheta}_{rcd}$ in Theorems~\ref{the:cef_con}-\ref{the:cefk}, respectively. All proofs of these theorems are given in Appendices B.1 and B.2. Without loss of generality, we assume $m = \min\{n_k\}, k = 1,\cdots, K$, throughout the rest of this paper, whenever applicable. 
\begin{theorem}
	\label{the:cef_con}
	Under regularity conditions (C1-C3) and (C4.0) given in Appendix A and the homogeneity of parameters $\btheta_{k0} = \btheta_0, k = 1,\cdots, K$,  meta estimator $\hat{\btheta}_{rcd}$ is consistent, namely,
	\[\hat{\btheta}_{rcd} \overset{p}{\rightarrow} \btheta_0, \hspace{0.5cm} \text{as} \hspace{0.2cm} m\to \infty.\]
\end{theorem}
\begin{theorem}
	\label{the:cef_norm}	
	Under the same conditions of Theorem~\ref{the:cef_con} and an additional condition (C4.1), meta estimator $\hat{\btheta}_{rcd}$ is asymptotically normally distributed, namely, $$\sqrt{n}\left(\hat{\btheta}_{rcd} - \btheta_0\right) \stackrel{d}{\rightarrow} \mathcal{N}\left(\bm{0}, \bj_{cd}^{-1}(\btheta_0)\right), \hspace{0.5cm} \text{as} \hspace{0.2cm} m \rightarrow \infty,$$
		where $\bj_{cd}(\btheta_0) = \lim_{m\to \infty}\left(\sum_{k=1}^{K}n_k\bj_{k}(\btheta_0)/n\right)$. 
\end{theorem}
Note that according to the definition of $m$, we have $n = \sum_{k=1}^{K}n_k > Km$. Thus $n \to \infty$ as $m \to \infty$. 
It follows from Theorem~\ref{the:cef_norm} that the convergence rate of $\hat{\btheta}_{rcd}$ is of order $n^{-1/2}$, not of order $m^{-1/2}$. This presents an important difference from subsampling strategy \cite[]{mahoney2011randomized, ma2015statistical}, in which the asymptotic convergence rate of their estimators is usually of an order given by the subsample size.

In practice, when the number of computing nodes in parallelization increases, {\em i.e.}, $K \to \infty$, the following Theorem~\ref{the:cefk} shows the asymptotic properties of the proposed Rao-CD meta estimator $\hat{\btheta}_{rcd}$. 
\begin{theorem}
	\label{the:cefk}
	For $K = O(n^{1/2 - \delta})$ with some positive constant $\delta < 1/2$,
	\begin{itemize}
		\item[(a)] under the same conditions of Theorem~\ref{the:cef_con}, the estimation consistency for $\hat{\btheta}_{rcd}$ given in Theorem~\ref{the:cef_con} remains true;
		\item[(b)] moreover, under the same conditions of Theorem~\ref{the:cef_norm}, the asymptotic normality in Theorem~\ref{the:cef_norm} holds for $\hat{\btheta}_{rcd}$ with the information matrix  $j_{cd}(\btheta_0) = \lim_{K \to \infty}\lim_{m\to \infty}\left(\sum_{k=1}^{K}n_kj_{k}(\btheta_0)/n\right)$.	
	\end{itemize}

\end{theorem}

\begin{remark}
In the case where each sub-dataset has the same size, i.e., $n_k \equiv m, k = 1, \dots, K$, it is easy to obtain that in Theorem~\ref{the:cefk} under $\delta<0.5$, together with Theorem~\ref{the:cef_con} and Theorem~\ref{the:cef_norm} under $\delta = 0.5$ or $K = O(1)$, $K = O\left(m^{(1- 2\delta)/(1 + 2\delta)} \right)$ for $\delta \in (0, 0.5]$ . \cite{lin2011aggregated} derived the asymptotic distribution for a quasi-likelihood estimator under the assumption that $K = O(m^{\gamma})$ for a positive constant $\gamma < \min\left\{1 - 2\alpha, 4\alpha -1 \right\}$ with $\alpha \in (1/4, 1/2)$, which is much narrower than the range given in Theorem~\ref{the:cefk}.  
\end{remark}

\subsection{Asymptotic efficiency}
In this section, we first present the asymptotic equivalency of the two types of CD estimators, $\hat{\btheta}_{rcd}$ and $\hat{\btheta}_{wcd}$, in Theorem~\ref{the:cd&cef}. This theorem provides the theoretical basis for a fast algorithm to implement $\hat{\btheta}_{rcd}$ in Section~\ref{sec:imp}. Then we discuss the issue between $\hat{\btheta}_{rcd}$ and $\hat{\btheta}_{full}$ in Theorem~\ref{the:corie}. The proofs of Theorems~\ref{the:cd&cef} and~\ref{the:corie} are given in Appendices B.4 and B.6, respectively.  
\begin{theorem}
	\label{the:cd&cef}
	If conditions (C1)-(C3) and (C4.2) hold, we have 
	$$\|\hat{\btheta}_{rcd} - \hat{\btheta}_{wcd}\|_2 = O_p(Kn^{-1}).$$
\end{theorem} 

\begin{remark}
	Conditions (C4.0), (C4.1) and (C4.2) are weaker conditions than the typical smoothness assumptions adopted in the theory of the estimating functions ({\em i.e.}, twice continuously differentiable). These conditions (C4.0), (C4.1) and (C4.2) have also been considered in \cite{pakes1989simulation}, \cite{newey1994large}, among others. In other words, these conditions automatically hold when estimating function $\psi$ is twice continuously differentiable. 
\end{remark}

\begin{remark}
	According to Theorem~\ref{the:cd&cef}, the asymptotic equivalency between $\hat{\btheta}_{rcd}$ and $\hat{\btheta}_{wcd}$ is accurate up to the second order $O_p(n^{-1})$ under fixed $K$. When $K$ increases, under the condition $K = O(n^{1/2 - \delta}), \delta \in (0, 0.5)$ in Theorem~\ref{the:cefk}, the resulting order of asymptotic equivalency becomes $O_p(n^{-1/2 - \delta})$, slightly slower than the rate $n^{-1}$. 
\end{remark}

Theorems \ref{the:cefk} and \ref{the:cd&cef} establish the estimation consistency and asymptotic normality of both Rao-CD and Wald-CD meta estimators as $K \to \infty$. These important theoretical properties are useful to implement these meta estimators in the MapReduce paradigm. With no surprise, the number of parallel datasets, $K$, cannot increases at an arbitrarily fast rate as the information attrition can affect the quality of estimation within each sub-dataset. Intuitively, ensuring the goodness of fit for individual estimator is of the first importance in order to yield a desirable meta estimator. Technically, it is attributed to the fact that the estimation bias for a sub-dataset is at order of $O_p(n^{-1}_k)$, which does not vanish over the data aggregation relatively to the variance of the resultant meta estimator. In other words, the proposed combination procedure helps improve the order of the variance to the parametric rate $O_p(n^{-1/2})$, whereas the order of the estimation bias remains the same at the rate of sub-dataset size. Consequently, an increase in the number of computing nodes $K$ should be controlled in such a way that the estimation bias is ignorable relative to the variance of the meta estimator. From a theoretical point of view, {one of the directions to improve is through de-biased methods \cite[{\em e.g.}, ][]{firth1993bias, cordeiro1991bias}, which may permit $K$ increases to infinity at a faster rate than what has been obtained in this paper.} However, from a practical point of view, allocating the number of CPUs is constrained by budget and available computing sources, and thus it is not necessary to let $K$ diverge at an arbitrary rate.

We now turn to the asymptotic efficiency of $\hat{\btheta}_{rcd}$ relative to that of $\hat{\btheta}_{full}$, which is the estimator obtained by processing the entire data once from the following estimating equation, where MapReduce strategy is not used; that is, $\hat{\btheta}_{full}$ satisfies 
\begin{eqnarray}
\label{met:de}
\psi_{full}(\bW; \hat{\btheta}_{full}) \overset{def}{=} n^{-1}\sum_{k=1}^{K}n_k\psi_{k\_sub}(\bW^{(k)}; \hat{\btheta}_{full}) = \bm{0}.
\end{eqnarray}  
The standard theory of estimation functions claims that under the same conditions of Theorem~\ref{the:cef_norm}, we have both estimation consistency and asymptotic normality, 
	$$\hat{\btheta}_{full}\overset{p}{\rightarrow} \btheta_0, \hspace{0.2cm}\mbox{and}\hspace{0.2cm} \sqrt{n}\left(\hat{\btheta}_{full} - \btheta_0\right) \stackrel{d}{\rightarrow} \mathcal{N}\left(\bm{0}, \bj^{-1}(\btheta_0)\right), \hspace{0.2cm} \text{as} \hspace{0.2cm} n \to \infty, $$
	where Godambe information $\bj(\btheta) = \bs^{T}(\btheta)\bv^{-1}(\btheta)\bs(\btheta)$, with sensitivity matrix $\bs(\btheta) = E\left\{\bS_{n}(\btheta)\right\}$, and variability matrix $\bv(\btheta) = \mbox{Var}\left\{\sqrt{n}\psi_{full}(\bW; \btheta)\right\}$.

It is interesting to note that the root of equation~(\ref{met:de}), $\hat{\btheta}_{full}$, may be regarded as a minimizer of the following quadratic estimation function: 
	\begin{eqnarray}
	\label{gmm:full}
		\hat{\btheta}_{full} = \arg\min_{\btheta}\left\{\bpsi_{n}(\bW; \btheta)^{T}\mathbb{S}^{-1}_{n}(\btheta)\bpsi_n(\bW; \btheta)\right\},		
	\end{eqnarray}
where $\mathbb{S}_{n}(\btheta) = \mbox{block-diag}\left\{\bS_{n_1}(\btheta), \dots, \bS_{n_K}(\btheta)\right\}$ and $\bpsi_n(\bW; \btheta)$ is the extended vector of estimating functions defined in Section~\ref{sec:com}. 
The method given in~\eqref{gmm:full} is quite similar to the so-called aggregated estimation equation (AEE) proposed by \cite{lin2011aggregated}; that is,  \[\hat{\btheta}_{AEE} = \arg\min_{\btheta}\left\{\bpsi_n(\bW; \btheta)^{T}\hat{\mathbb{S}}^{-1}_n\bpsi_n(\bW; \btheta) \right\},\] where $\hat{\mathbb{S}}_{n} = \mbox{block-diag}\left\{\hat{\bS}_{n_1}, \dots, \hat{\bS}_{n_K}\right\}$ is a consistent estimator of $\mathbb{S}_{n}(\btheta)$. Hence, $\hat{\btheta}_{AEE}$ and $\hat{\btheta}_{full}$ may be different numerically under finite samples, but they have the same asymptotic distribution. 

We gain two important insights by comparing expressions~(\ref{gmm:cef}) and (\ref{gmm:full}) in terms of the two types of weighting matrices, $\hat{\mathbb{V}}^{-1}_{n}$ versus $\hat{\mathbb{S}}^{-1}_{n}$, leading to $\hat{\btheta}_{rcd}$ (asymptotically equivalent to $\hat{\btheta}_{wcd}$) and $\hat{\btheta}_{AEE}$ (asymptotically equivalent to $\hat{\btheta}_{full}$), respectively. Note that $\mathbb{V}_n$ and $\mathbb{S}_n$ are different in general in the context of estimating functions. Because of such weighting differences, according to Hansen's theory of GMM, $\hat{\btheta}_{rcd}$ (or $\hat{\btheta}_{wcd}$) will be asymptotically more efficient than $\hat{\btheta}_{AEE}$ (or $\hat{\btheta}_{full}$). This insight is summarized in Theorem~\ref{the:corie} below.

\begin{theorem}
	\label{the:corie} 
	Under the same conditions of Theorem~\ref{the:cef_norm}, we have the following inequality of Godambe information, 
	$$j_{cd}(\btheta_0) \geq j(\btheta_0),$$
	where the equality holds if and only if the Bartlett identity holds for estimating function $\psi(\cdot)$({\em e.g.} $\psi$ being the score function), $\bs_{k}(\btheta_0) \equiv \bv_k(\btheta_0), k = 1, \dots, K$; or there exists a homogeneous asymptotic godambe information across all $K$ sub-datasets, i.e., $\bj_{1}(\btheta_0) \equiv \bj_2(\btheta_0) \equiv \cdots \equiv \bj_{K}(\btheta_0)$. 
\end{theorem}

Theorem~\ref{the:corie} indicates the Rao-CD meta estimator $\hat{\btheta}_{rcd}$ is asymptotically at least as efficient as the one-time estimator, $\hat{\btheta}_{full}$, when the same estimating method is applied with individual sub-datasets in the MapReduce paradigm and with the entire data once in a ``God-made" computer. If the homogeneity of information matrices across individual sub-datasets is violated, $\hat{\btheta}_{rcd}$ will produce better efficiency than the $\hat{\btheta}_{full}$. The result is somewhat counter-intuitive; but it always occurs in actual data analysis, where with finite samples one would yield unequal empirical information matrices $\hat{\bV}_{n_k}$ and $\hat{\bS}_{n_k}$. This efficiency improvement is actually rooted in the fact that the way of weighting in~(\ref{gmm:cef}) is optimal \cite[]{hansen1982large},  and thus, better than that in~(\ref{gmm:full}). Similar results are also found in \cite{zeng2015ran} under random effects models, and in \cite{hu2000estimating}, who pointed out that the studentized estimating function bootstrap is second-order accurate in comparison to the first order approximation of the estimating function bootstrap. 
When the estimating function is the score function, the same result has already been found in \cite{lin2010relative} and \cite{liu2015multivariate} for the maximum likelihood estimation. 

Another important property of the Rao-CD meta estimation concerns estimation robustness against data contamination. Note that the variance-based weighting scheme in~(\ref{gmm:cef}) creates an automatic down-weighting for any data cases associated with large residual values in the estimation procedure; see \cite{qu2004assessing, preisser1999robust, hampel2011robust}, among others. This down-weighting mechanism makes the Rao-CD meta estimation more robust than the one-time estimator, $\hat{\btheta}_{full}$, obtained from \eqref{met:de} or \eqref{gmm:full}. For the Wald-CD meta estimator in~(\ref{gmm:cd}), this weighting scheme takes approximately a form of the variance of $(\hat{\btheta}_k - \btheta)^{T}\hat{\bS}_{n_k}$, hence the robustness of the Rao-CD meta estimator is also shared with the Wald-CD meta estimator. In addition to the automatic down-weighting scheme in the Rao-CD meta estimation approach, data split actually allocates outliers into some of the sub-datasets, affecting potentially a few $\hat{\btheta}_k$'s in those sub-datasets that contain outliers. This dilution of influential cases via the data division adds another layer of protection for the Rao-CD meta estimation in addition to the down-weighting mechanism, which ensures greater robustness of the CD-based estimation and inference against contaminated data cases. In section~\ref{sec:simu}, we will use simulation study to illustrate the robustness of the Rao-CD meta estimation approach. 

\section{Implementation}
\label{sec:imp}

It follows immediately from the definition of the Wald-CD meta estimator in~(\ref{gmm:cd}) that $\hat{\btheta}_{wcd}$ at the Reduce-step based on $K$ mapped sub-datasets is given by  
\begin{eqnarray}
\label{eq:mapcd}
\hat{\btheta}_{wcd} = \left\{\sum_{k=1}^{K}n_k\hat{\bS}^{T}_{n_k}\hat{\bV}^{-1}_{n_k}\hat{\bS}_{n_k} \right\}^{-1}\left\{\sum_{k=1}^{K}n_k\hat{\bS}^{T}_{n_k}\hat{\bV}^{-1}_{n_k}\hat{\bS}_{n_k}\hat{\btheta}_k\right\}.
\end{eqnarray}
This closed-form expression of $\hat{\btheta}_{wcd}$ in~(\ref{eq:mapcd}) only involves summary statistics $(\hat{\bS}_{n_k}, \hat{\bV}^{-1}_{n_k}, \hat{\btheta}_k), k = 1, \dots, K$, that are calculated separately in the Map-step on individual computing nodes. Apparently, \eqref{eq:mapcd} presents a scalable parallel calculation with no need of loading the entire data into a common server, and thus
reduces considerable amount of computation time. Note that the AEE estimator $\hat{\btheta}_{AEE}$ can also be implemented similarly in this scalable MapReduce framework.

The proposed Rao-CD meta estimation in~(\ref{eq:cef}) may be implemented by the Newton-Raphson iterative algorithm. To do so, taking the second-order Taylor expansion of~(\ref{eq:cef}) around the $\hat{\btheta}_{wcd}$, we have 
\begin{eqnarray*}
&&\hat{\btheta}_{rcd} \approx \hat{\btheta}_{wcd} + \left\{\sum_{k=1}^Kn_k\bS^{T}_{n_k}(\hat{\btheta}_{wcd})\hat{\bV}^{-1}_{n_k}\bS_{n_k}(\hat{\btheta}_{wcd})\right\}^{-1}\\
&&\hspace{2.5cm} \times 
\left[\sum_{k=1}^Kn_k\bS^{T}_{n_k}(\hat{\btheta}_{wcd})\hat{\bV}^{-1}_{n_k}\left\{\psi_{k\_sub}(\bW^{(k)}; \hat{\btheta}_{wcd}) - \psi_{k\_sub}(\bW^{(k)}; \hat{\btheta}_{k})\right\}\right] \nonumber\\
&&\approx \hat{\btheta}_{wcd} - \left\{\sum_{k=1}^Kn_k\bS^{T}_{n_k}(\hat{\btheta}_{wcd})\hat{\bV}^{-1}_{n_k}\bS_{n_k}(\hat{\btheta}_{wcd})\right\}^{-1}
\left\{\sum_{k=1}^Kn_k\bS^{T}_{n_k}(\hat{\btheta}_{k})\hat{\bV}^{-1}_{n_k}\bS_{n_k}(\hat{\btheta}_{k})\left(\hat{\btheta}_{wcd} - \hat{\btheta}_k\right)\right\}.\nonumber
\end{eqnarray*}
It is interesting to note that the second factor in the second term of the above expression equals to zero because the expression of $\hat{\btheta}_{wcd}$ in~(\ref{eq:mapcd}). These two types of CD meta estimators are numerically very close to each other, where $\hat{\btheta}_{rcd}$ may be regarded as a solution obtained by a one-step Newton-Raphson update from the $\hat{\btheta}_{wcd}$ in~\eqref{eq:mapcd}. The associated approximation error between them is explicitly gauged in Theorem~\ref{the:cd&cef}, with the theoretical order of error being $O_p(Kn^{-1})$, which supports the above numerical approximation. This point of view is particularly appealing for big data computation with large $n$.

To establish statistical inference, we propose to estimate the variance of $\hat{\btheta}_{rcd}$ by its empirical asymptotic variance, namely, $\hat{\text{Var}}(\hat{\btheta}_{rcd}) = \bJ^{-1}_n(\hat{\btheta}_{rcd})$, with $\bJ_{n}(\hat{\btheta}_{rcd}) \overset{def}{=} n^{-1}\sum_{k=1}^Kn_k\bJ_{n_k}(\hat{\btheta}_{rcd})$,
where $\bJ_{n_k}(\btheta) = \bS^{T}_{n_k}(\btheta)\bV^{-1}_{n_k}(\btheta)\bS_{n_k}(\btheta)$. In the MapReduce paradigm, to avoid using the entire data, we propose to approximate $\bJ_n(\hat{\btheta}_{rcd})$ by the following estimate: 
\begin{eqnarray}
\label{eq:estv}
\bJ_n^{a}(\hat{\btheta}_1, \dots, \hat{\btheta}_{K}) = n^{-1}\sum_{k=1}^{K}n_k\bJ_{n_k}(\hat{\btheta}_k),
\end{eqnarray} 
where $\bJ_{n_k}(\hat{\btheta}_k)$ is the empirical Godambe information matrix of $\hat{\btheta}_k$ obtained in the Map-step on a single computing node. The following theorem provides a theoretical assessment of the approximation error incurred by the estimate in~\eqref{eq:estv}.

\begin{theorem}
	\label{the:var}
	Under conditions (C1)-(C3) and (C4.1), we have 
	\begin{eqnarray}
	\bJ_n^{a}(\hat{\btheta}_1, \dots, \hat{\btheta}_{K}) = \bj_{cd}(\btheta_0) + O_p(n^{-1/2} + Kn^{-1}).
	\end{eqnarray}	
\end{theorem}	 

The proof of Theorem~\ref{the:var} is given in Appendix B.5. Theorems~\ref{the:cd&cef} and \ref{the:var} suggest that for big data computation with very large $n$, the two types of meta CD estimators $\hat{\btheta}_{rcd}$ and $\hat{\btheta}_{wcd}$ are numerically very close to each other in terms of both point estimation and statistical inference. However, it is worth pointing out that the Rao-CD approach presents a much more appealing theoretical framework to establish and interpret theoretical properties, and more importantly, carry out relevant analytic justification. We have implement the proposed Rao-CD and Wald-CD meta estimation methods in the Hadoop programming framework using Python language. The code has been applied to conduct simulation studies and real data analyses. The software is available for download at http://www.umich.edu/$\sim$songlab/software.html\#RCD.

\section{Several examples}
\label{sec:ex}
The Rao-CD approach is applicable to a wide range of important statistical models. Here
we present three representatives that are of great popularity in practice.

\subsection{Quantile regression}
Denote regression data by $\bW = (Y, \bX)$, where $Y$ is the outcome and $\bX$ is a vector of regressors. Let the conditional distribution function of variable $Y$ given $\bX$ be $F_{Y\mid \bX}(y)$, and let the $\tau$th quantile of $Y\mid \bX$ be 
$Q_{Y\mid \bX}(\tau) = F^{-1}_{Y\mid \bX}(\tau) = \inf\{y: F_{Y\mid \bX}(y) \geq \tau\}$, $\tau \in (0, 1)$. According to \cite{koenker2005quantile}, a quantile regression model takes a form of $$Q_{Y\mid \bX}(\tau) = \bX^{T}\btheta_0.$$

Now applying the MapReduce paradigm, one may divide the data into K sub-datasets and use the following estimating function to estimate parameter $\btheta$ with the $k^{th}$ sub-dataset at one computing node:
\begin{eqnarray*}
	\psi_{k\_{sub}}(\bW^{(k)}; \btheta) = n^{-1}_k\sum_{i=1}^{n_k}\bX_{k,i}\left\{I(y_{k,i} - \bX^{T}_{k,i}\btheta \leq 0) - \tau\right\}, \hspace{0.2cm} k = 1, \dots, K,
\end{eqnarray*}
where $I(\cdot)$ is the indicator function. For a solution of the above equation $\hat{\btheta}_k$ satisfying $\psi_{k\_sub}(\bW^{(k)}; \hat{\btheta}_k) = \bm{0}$, the standard theory of quantile regression  \cite[]{koenker2005quantile} establishes the following asymptotic distributions, under some regularity conditions, 
\begin{eqnarray*}
\sqrt{n_k}\psi_{k\_{sub}}(\bW^{(k)}; \btheta_0) &\stackrel{d}{\rightarrow}& \mathcal{N}\left(
\bm{0}, \tau(1 - \tau) \bv_{k}(\btheta_0)
\right), \hspace{0.5cm} \text{as} \hspace{0.2cm} n_k \to \infty, \nonumber\\
\sqrt{n_k}\left\{\hat{\btheta}_k - \btheta_0\right\} &\stackrel{d}{\rightarrow}& \mathcal{N}\left(\bm{0}, \tau(1-\tau)\left\{\bs^{-1}_{k}(\btheta_0)\right\}^{T}\bv_{k}(\btheta_0)\bs_{k}^{-1}(\btheta_0)\right),
\hspace{0.5cm} \text{as} \hspace{0.2cm} n_k \to \infty,
\end{eqnarray*}
where $\bv_{k}(\btheta_0) = E_{\btheta_0}(\bX_{k,i}\bX_{k,i}^{T})$, and 
$\bs_{k}(\btheta_0) = E_{\btheta_0}\left\{\bX_{k, i}\bX_{k, i}^{T}f_{Y\mid \bX}(\bX_{k, i}^{T}\btheta_0)\right\}$. 
It follows from~(\ref{gmm:cef}) that the Rao-CD meta estimator is given by: 
\begin{eqnarray}
\label{gmm_cef:quan}
\hat{\btheta}_{rcd} = \arg\min_{\btheta}\sum_{k=1}^{K}n_k\psi^{T}_{k\_{sub}}(\bW^{(k)}; \btheta)\hat{\bV}^{-1}_{n_k}\psi_{k\_{sub}}(\bW^{(k)}; \btheta),
\end{eqnarray}
with $\hat{\bV}_{n_k} = n^{-1}_k\sum_{i=1}^{n_k}\bX_{k, i}\bX_{k, i}^{T}$, independent of $\hat{\btheta}_k$. Also, by~({\ref{gmm:cd}}), the Wald-CD meta estimator is obtained as,
$$\hat{\btheta}_{wcd} = \arg\min_{\btheta}\sum_{k=1}^{K}n_k(\hat{\btheta}_k - \btheta)^{T}\hat{\bS}^{T}_{n_k}\hat{\bV}^{-1}_{n_k}\hat{\bS}_{n_k}(\hat{\btheta}_k - \btheta),
$$
where the empirical sensitivity matrix $\bS_{n_k}(\btheta)$ involves estimation of unknown density $f_{Y\mid \bX}$. Estimating $f_{Y \mid \bX}(\cdot)$ may be tedious and unstable when $n_k$ is not large. From this perspective, the Rao-CD meta estimator $\hat{\btheta}_{rcd}$ may be numerically more stable than the Wald-CD meta estimator $\hat{\btheta}_{wcd}$ in cases where the density $f_{Y\mid \bX}$ is hard to estimate. Although directly minimizing the quadratic term (\ref{gmm_cef:quan}) is in favor of numerical stability, it is prohibited in the MapReduce framework as the direct optimization in~(\ref{gmm_cef:quan}) requires reloading the entire data. Alternatively, the Apache Spark platform \cite[]{zaharia2010spark} may be used to overcome the challenge of implementation as Spark provides a more flexible management of data reloading. This is beyond the scope of this paper. When the size of each sub-dataset is set large enough under which the density $f_{Y\mid \bX}$ is well estimated, we may use the one-step updating strategy given in Section~\ref{sec:imp}, where a completely parallelized calculation gives rise to a fast and simple implementation. Some numerical results are shown in Section~\ref{sec:simu} for the advantage of this parallelized computing scheme.  

\subsection{Generalized estimation equation}
Consider longitudinal data $\bW = \left\{\bW_i = (\by_i, \bx_i), i = 1,\cdots, n\right\}$ consisting of $n$ independent realizations $\bW_i, i = 1, \dots, n$, with subject $i$ being observed repeatedly at $l_i$ times. In the literature of longitudinal analysis, generalized estimating equation (GEE), proposed by \cite{liang1986longitudinal}, is one of the most widely used methods, which is a quasi-likelihood approach based only on the first two moments of the data distribution. Denote the first two conditional moments of $\bY$ given $\bX = \bx$ by
$E(\bY\mid \bX = \bx_i) = \mu(\bx_i; \btheta) = \left[g(\bx^T_{i, 1}\btheta), \dots, g(\bx^{T}_{i, l_i}\btheta)\right]^{T}$, and $cov(\bY\mid \bX = \bx_i) = \sigma^2 \Sigma_i(\btheta, \brho) = 
\sigma^2 G(\bx_i; \btheta)^{1/2}R(\brho)G(\bx_i; \btheta)^{1/2}$, where $g(\cdot)$ is a known link function, $\sigma^2 > 0$ is the dispersion parameter, $G(\bx_i; \btheta) = \text{diag}\left[V\{g(\bx^{T}_{i, 1}\btheta)\}, \dots, V\{g(\bx^{T}_{i, l_i}\btheta)\}\right]$ is a diagonal matrix with $V(\cdot)$ being a known variance function, and $R(\brho)$ is a working correlation matrix, which is fully characterized by a correlation parameter vector $\brho$. In the MapReduce paradigm, to estimate the parameter of interest, $\btheta$, the following GEE for the $k^{th}$ sub-dataset is used:
\begin{eqnarray}
\label{est:lon}
\psi_{k\_{sub}}(\bW^{(k)}; \btheta, \brho) = n_k^{-1}\sum_{i=1}^{n_k}\sigma^{-2}\bx_{k, i}D_{k, i}(\btheta)\Sigma^{-1}_{k, i}(\btheta, \brho)\left\{\by_{k, i} - \mu(\bx_{k, i}; \btheta)\right\} = \bm{0},
\end{eqnarray}
where $\bx_{k, i} = (\bx_{k, i1}, \bx_{k, i2}, \dots, \bx_{k, il_i})^{T}$, $D_{k, i}(\btheta) = \mbox{diag}\left\{\dot{g}(\bx_{k, i1}^{T}\btheta), \dots, \dot{g}(\bx_{k, il_i}^{T}\btheta)\right\}$, and $\Sigma_{k, i}(\btheta, \brho) = G_{k, i}(\btheta)^{1/2}R_{k, i}(\brho)G_{k, i}(\btheta)^{1/2}$,  with 
$G_{k, i}(\btheta) = \mbox{diag}\left[V\{g(\bx_{k, i1}^{T}\btheta)\}\right.$
$\left., \dots, V\{g(\bx_{k, il_i}^{T}\btheta)\}\right]$. In addition, the empirical sensitivity and variability matrices are given by 
 \begin{eqnarray*}
 	\bS_{n_k}(\btheta; \brho) &=& n^{-1}_k\sum_{i = 1}^{n_k}\bx_{k, i}D_{k, i}(\btheta)\Sigma_{k, i}^{-1}(\btheta, \brho)D_{k,i}(\btheta)\bx_{k, i}^{T},\\
 	\bV_{n_k}(\btheta; \brho) &=& n^{-1}_k\sum_{i = 1}^{n_k}\bx_{k, i}D_{k, i}(\btheta)\Sigma_{k, i}^{-1}(\btheta, \brho)\left\{y_{k, i} - \mu(\bx_{k, i}; \btheta)\right\}\left\{y_{k, i} - \mu(\bx_{k, i}; \btheta)\right\}^{T}\Sigma_{k, i}^{-1}(\btheta, \brho)D_{k, i}(\btheta)\bx_{k, i}^{T}.
 \end{eqnarray*}
Note that in the GEE, both information matrices above are easy to be evaluated numerically, the implementation in the Reduce-step proposed in Section~\ref{sec:imp} is straightforward through the following one-step updating procedure: 
$$ \hat{\btheta}_{rcd} = \left\{\sum_{k=1}^{K}n_k\bS^{T}_{n_k}(\hat{\btheta}_k; \hat{\brho}_k)\bV^{-1}_{n_k}(\hat{\btheta}_k; \hat{\brho}_k)\bS_{n_k}(\hat{\btheta}_k; \hat{\brho}_k)\right\}^{-1}\left\{ \sum_{k=1}^{K}n_k\bS^{T}_{n_k}(\hat{\btheta}_k; \hat{\brho}_k)\bV^{-1}_{n_k}(\hat{\btheta}_k; \hat{\brho}_k)\bS_{n_k}(\hat{\btheta}_k; \hat{\brho}_k)\hat{\btheta}_k \right\},
$$
where $\hat{\btheta}_k$ and $\hat{\brho}_k$ are obtained by the standard GEE software, such as the R package \verb|geepack| and the Python package \verb|Statsmodels|, at each individual computing node in the operation of the Map-step.  

Note that in the above calculation, the nuisance parameter $\brho$ is estimated separately with respective to sub-datasets, $\hat{\brho}_k, k = 1, \dots, K$. In other words, even if a common working correlation structure $R(\brho)$ is assumed for the entire data, the implementation by the MapReduce scheme gives rise to 
heterogeneous estimates of the correlation structure. This implies higher variation in $\bV_{n_k}(\hat{\btheta}_k; \hat{\brho}_k)$, leading to a stronger locally varying weighting scheme across sub-datasets. Consequently, from the point view of GMM, the objective function of the CD meta estimation method \eqref{gmm:cef} appears to have lower variability than the full data based objective function in \eqref{gmm:full}. Similar findings are reported in the literature of the inverse probability weighting (IPW) method for missing data analysis, where the weight matrix may be estimated nonparametrically (NPIPW), or parametrically (PIPW), or being fixed (FIPW). According to \cite{chen2015efficient}, the estimator from NPIPW has shown to have the least variance among the three methods. 

\subsection{Survival data analysis}
In the Cox proportional hazards model \citep{cox1972regression, cox1975partial}, the hazard function $\lambda(t)$ is specified as $\lambda(t) = \lambda_0(t)\exp(\bX^{T}\btheta)$, where $\lambda_0(t)$ is an unknown baseline hazard function. The method of the partial likelihood is known to provide an efficient estimation of regression parameter, $\btheta$. With the $k$th sub-dataset $\bW^{(k)} = \{(\delta_{k, i}, \mathcal{T}_{k, i}, \bX_{k, i}): i = 1,\dots, n_k\}$, the partial score for $\btheta$ is given as follows:
\begin{eqnarray*}
	\psi_{k\_{sub}}(\bW^{(k)}; \btheta) = n^{-1}_k\sum_{i=1}^{n_k} \delta_{k, i}\left\{\bx_{k, i} - \frac{\sum_{j=1}^{n_k}I(\mathcal{T}_{k, j} \geq \mathcal{T}_{k, i})\exp(\bx_{k, j}^{T}\btheta)\bx_{k, j}}
	{\sum_{j=1}^{n_k}I(\mathcal{T}_{k, j} \geq \mathcal{T}_{k, i})\exp(\bx_{k, j}^{T}\btheta)}\right\},
\end{eqnarray*}
where $\delta_{k, i}$ is the indicator of failure time $T_{k, i}$ being observed ($\delta_{k, i} = 1$) or censored ($\delta_{k, i} = 0$), and $\mathcal{T}_{k, i} = \min(T_{k, i}, C_{k, i})$ with $C_{k, i}$ being the censoring time. Denote $\hat{\btheta}_k$ as the partial likelihood estimator that satisfies $\psi_{k\_{sub}}(\bW^{(k)}; \hat{\btheta}_k) = \bm{0}, k = 1, \dots, K.$ By~(\ref{gmm:cef}) and (\ref{gmm:cd}), the Rao-CD and Wald-CD meta estimators $\hat\btheta_{rcd}$ and $\hat{\btheta}_{wcd}$ are, respectively, the solutions of the following estimating equations:
\begin{eqnarray*}
	&&\Psi_{R}(\btheta) = n^{-1/2}\sum_{k=1}^{K}n_k\bS^{T}_{n_k}(\btheta)\hat{\bV}^{-1}_{n_k}\psi_{k\_{sub}}(\bW^{(k)}; \btheta) = \bm{0}, \nonumber\\
	&&\Psi_{W}(\btheta) = n^{-1/2}\sum_{k=1}^{K}n_k\hat{\bS}^{T}_{n_k}\hat{\bV}^{-1}_{n_k}\hat{\bS}_{n_k}(\hat{\btheta}_k - \btheta) = \bm{0},
\end{eqnarray*}
where $\bS_{n_k}(\btheta) = -\dot{\psi}_{k\_{sub}}(\btheta)$ and $\bV_{n_k}(\btheta) = n^{-1}_k\sum_{i=1}^{n_k} \delta_{k, i}\left\{\bx_{k, i} - \frac{\sum_{j=1}^{n_k}I(\mathcal{T}_{k, j} \geq \mathcal{T}_{k, i})\exp(\bx_{k, j}^{T}\btheta)\bx_{k, j}}
{\sum_{j=1}^{n_k}I(\mathcal{T}_{k, j} \geq \mathcal{T}_{k, i})\exp(\bx_{k, j}^{T}\btheta)}\right\}^{\otimes 2}$, and $\hat{\bV}_{n_k} = \bV_{n_k}(\hat{\btheta}_k)$. Since the estimating function $\Psi_{R}(\btheta)$ above has continuous second-order derivatives with respect to $\btheta$, $\hat{\btheta}_{rcd}$ and $\hat{\btheta}_{wcd}$ are close to each other at an order of $O_p(Kn^{-1})$, making the implementation of the one-step updating scheme for $\hat{\btheta}_{rcd}$ very straightforward. Numerically, when $n$ is large, the difference between $\hat{\btheta}_{rcd}$ and $\hat{\btheta}_{wcd}$ is ignorable.

One possible technical issue for fitting the Cox model in the MapReduce paradigm pertains to the baseline hazard function $\lambda_0(t)$. Though $\lambda_0(t)$ is not necessary to be estimated in the above partial likelihood method with individual sub-datasets, the implicit assumption concerning the actual parameter space differs between the parallel CD estimation $\hat{\btheta}_{rcd}$ and the full data estimation $\hat{\btheta}_{full}$. The former assumes different baseline hazard, $\lambda_{0, k}(t), k = 1, \dots, K$, whereas the latter assumes a common baseline hazard $\lambda_0(t)$. This implies that the non-parallel full-data estimator $\hat{\btheta}_{full}$ is subject to more restrictions regarding the homogeneity on the baseline hazard function in comparison to the parallel CD approach. Some consequences of this difference include: (a) the CD estimator $\hat{\btheta}_{rcd}$ tends to produce a slightly larger estimation variance compared to that of the full-data estimator $\hat{\btheta}_{full}$; and (b) the full-data estimator $\hat{\btheta}_{full}$ would be biased if the actual baseline functions were different across sub-datasets, for example, in multiple cohort studies. In addition, according to Theorem~\ref{the:corie}, when individual sub-datasets share the same baseline hazard, {\em say}, $\lambda_0(t)$, the Rao-CD meta estimation also provides an asymptotically efficient estimator of $\btheta$.

\section{Simulation experiments}
\label{sec:simu}
We now conduct simulation experiments to assess the performance of the proposed CD meta estimation method in the following aspects: (i) validity of inference in connection to Theorem~\ref{the:cef_norm} and Theorem~\ref{the:var}; (ii) scalability via parallel computation; (iii) efficiency in connection to Theorem~\ref{the:corie}; (iv) robustness against contaminated data and/or certain model heterogeneity; and (v) computational stability in connection to the one-step updating procedure given in Section~\ref{sec:imp}. We consider three classes of models, including quantile regression model, longitudinal GEE model, and Cox proportional hazards model. To avoid redundancy, we focus our evaluations on some of these five domains in each of these models, in order to control the length of this section.  

\subsection{Quantile regression}
In this subsection, the evaluation concerns three aspects (i), (iii) and (v). For (i), the validity of inference is examined in connection to the asymptotics of Theorem~\ref{the:cef_norm} and the estimation of the meta variance in Theorem~\ref{the:var}. For (iii), we further examine statistical power by comparing the CD estimators $\hat{\btheta}_{rcd}$ and $\hat{\btheta}_{wcd}$ with $\hat{\btheta}_{full}$. For (v), the assessment is focused on the one-step updating procedure in Section 5 as to whether the approximation error affects inference or not. Data are simulated from the following settings. Covariates $\bX$, consisting of $\bX_1, \dots, \bX_9$, are generated from a multivariate normal distribution with mean $\bm{0}$, identical marginal variance $\bm{1}$, and the compound symmetric correlation structure with $\rho = 0.5$. The response $Y_i$ is generated from a linear model $Y_i = \theta_0 + \bX_i^{T}\btheta + \varepsilon_{i}$, with $\theta_0 = 1$, $\btheta = (\theta_1, \dots, \theta_9)^{T} = (1, \cdots, 1)^{T}$ and $\varepsilon_i \stackrel{i.i.d.}{\sim} N(0, 1), i = 1, \dots, n$. Clearly, for any given percentile $\tau$, the conditional quantile is given by $Q_{Y\mid \bX}(\tau) = \btheta_0 + \Phi^{-1}(\tau) + \bX^{T}\btheta$, where $\Phi(\cdot)$ is the standard normal cumulative distribution function with mean 0 and variance 1 and $\Phi^{-1}(\cdot)$ is its quantile function. Here, we set $\tau = 0.5$ ({\em i.e.} median). We consider meta data consisting of $K = 20$ sub-datasets with an equal data size of $m = 500, 1\hspace{0.05cm}000, 2\hspace{0.05cm}000, 5\hspace{0.05cm}000$. In the evaluation, we calculate the absolute value of bias (ABIAS), empirical standard error over 500 replications (ESE), average asymptotic standard error (ASE), and $95\%$ coverage probability (CP) based on relevant respective asymptotic formulas. The simulation results are summarized in Table~\ref{tab:qu_robust} over 500 rounds of simulations, where three regression coefficients $\theta_4$, $\theta_6$ and $\theta_9$ of the nine coefficients are arbitrarily chosen to be included due to the limitation of space. For the other 6 parameters, the results are similar. In addition, we calculate the average relative efficiency (ARE) as well as the proportion of the relative efficiency (PRE) less than 1 for the meta estimators $\hat{\btheta}_{rcd}$ or $\hat{\btheta}_{wcd}$ with respect to $\hat{\btheta}_{full}$. Note that when two types of methods are equally efficient, the ARE and the ``PRE $<$ 1" should be 1 and $50\%$, respectively.

For the assessment of inference validity, it is evident that the ASE and the corresponding ESE are very comparable in all cases for the Rao-CD estimation, and thus the use of the asymptotic covariance matrix derived in Theorem~\ref{the:cef_norm} and its implementation in Theorem~\ref{the:var} are reasonable, at least in the current simulation model. To evaluate computation stability for the one-step updating scheme, from Table~\ref{tab:qu_robust} we see that as $m$ increases up to $2\hspace{0.05cm}000$ or larger, the coverage probabilities given by both Rao-CD and Wald-CD methods become very close to each other, both being near the nominal level of $95\%$. This confirms that the approximation error vanishes as $n \to \infty$, as shown in Theorem~\ref{the:cd&cef}. However, when the size of sub-dataset is $1\hspace{0.05cm}000$ or less, the coverage probability by the Wald-CD is worse than that by the Rao-CD, the former being more distance below the nominal level $95\%$. These numerical results indicate that the Rao-CD approach appears more reliable than the Wald-CD approach when the individual sub-dataset size is not large. To examine relative efficiencies, we see that both AREs of $\hat{\btheta}_{rcd}$ and $\hat{\btheta}_{wcd}$ to $\hat{\btheta}_{full}$ increase from 0.89 to 0.98 as $m$ increases, and the values of ``PRE $<$ 1" decrease from 99.6\% to 88\%. Clearly, Table~\ref{tab:qu_robust} suggests that the meta estimators $\hat{\btheta}_{rcd}$ and $\hat{\btheta}_{wcd}$ are more efficient than $\hat{\btheta}_{full}$, and that as $m\to \infty$, the difference tends to disappear, confirming the result of Theorem~\ref{the:corie}. 

\begin{table}
	\hspace{-2cm}\caption{\label{tab:qu_robust} The ABIAS, ESE, ASE, and CP of three chosen regression coefficients $\theta_{j}, j = 4, 6, 9$ for the quantile regression model at the quantile level $\tau = 0.5$, where Rao-CD $\hat{\btheta}_{rcd}$, Wald-CD $\hat{\btheta}_{wcd}$, FULL $\hat{\btheta}_{full}$ are compared with the sub-dataset size $m = 500, 1000, 2000, 5000$.
		The average relative efficiency (ARE) and the proportion of relative efficiency less than 1 (PRE $<$ 1) of Rao-CD $\hat{\btheta}_{rcd}$ and Wald-CD $\hat{\btheta}_{wcd}$ to FULL $\hat{\btheta}_{full}$ are listed.}\\
	\hspace{-1cm}\fbox{%
		\begin{tabular}{cl*{11}{c}}
			&&\multicolumn{3}{c}{$\theta_4$}&\multicolumn{3}{c}{$\theta_6$}&\multicolumn{3}{c}{$\theta_9$}\\
			$m$&&Rao-CD&Wald-CD&FULL&Rao-CD&Wald-CD&FULL&Rao-CD&Wald-CD&FULL\\
			\hline
%
	   $500$
			&ABIAS  &0.014 & 0.015  &0.013   &0.013&0.015&0.013  &0.014&0.015&0.013\\
			&ESE    &0.017 & 0.020  &0.017   &0.017&0.019&0.016  &0.017&0.020&0.017\\
			&ASE    &0.015 & 0.015  &0.017   &0.015&0.015&0.017  &0.015&0.015&0.017\\
			&CP    &0.914 &0.864   &0.946   &0.920&0.890&0.968  &0.924&0.866&0.950\\
			
			&ARE  &0.897 &0.897   &1.000   &0.893&0.893&1.000  &0.894&0.894&1.000\\
			&PRE $<$ 1   &99.6  &99.6    &--      &99.6 &99.6 &--     &99.6 &99.6 &--\\
   	  $1000$
			&ABIAS  &0.010   &0.010   &0.010    &0.010&0.010   &0.009   &0.010&0.010&0.009\\
			&ESE    &0.013   &0.013   & 0.012   &0.012& 0.013  &0.012   &0.012&0.012&0.012\\
			&ASE    &0.011   &0.011   & 0.012   &0.011& 0.011  &0.012   &0.011&0.011&0.012\\
		    &CP    &0.908   &0.914   & 0.930   &0.926& 0.928  &0.948   &0.934&0.928&0.962\\
		    
		    &ARE  &0.944   &0.944   &1.000    &0.944& 0.944  &1.000      &0.944&0.944&1.000\\
		    &PRE $<$ 1   &98.8    &98.8    &--       &98.2 & 98.2   &--      &98.4 &98.4 &--\\
  	  $2000$
  	  &ABIAS  &0.007   &0.007   &0.006    &0.007&0.007   &0.007   &0.007&0.007&0.007\\
  	  &ESE    &0.008   &0.008   &0.008    &0.009&0.008   &0.008   &0.009&0.009&0.008\\
  	  &ASE    &0.008   &0.008   &0.008    &0.008&0.008   &0.008   &0.008&0.008&0.008\\
  	  &CP     &0.948   &0.944   &0.962    &0.938&0.936   &0.956   &0.938&0.932&0.956\\
  	  
  	  &ARE    &0.969   &0.969   &1.000    &0.969&0.969   &1.000   &0.969&0.969&1.000\\
  	  &PRE $<$ 1   &96.6    &96.6    &--       &95.6 &95.6    &--      &96.6 &96.6 &--\\
  	  
  	  $5000$
  	  &ABIAS  &0.004   &0.004   &0.004    &0.004&0.004  &0.004   &0.004&0.004&0.004\\
  	  &ESE    &0.006   &0.005   &0.005    &0.005&0.005  &0.005   &0.005&0.005&0.005\\
  	  &ASE    &0.005   &0.005   &0.005    &0.005&0.005  &0.005   &0.005&0.005&0.005\\
  	  &CP     &0.934   &0.938   &0.950    &0.952&0.962  &0.964   &0.948&0.956&0.958\\
  	  
  	  &ARE    &0.984   &0.984   &1.000    &0.984&0.984  &1.000   &0.984&0.984&1.000\\
  	  &PRE $<$ 1   &88.4    &88.4    &--       &87.6 &87.6   &--      &88.8 &88.8 &--\\

		\hline\hline
		\end{tabular}}	
	\end{table}

\subsection{Longitudinal GEE model}
This subsection focuses on a comprehensive evaluation of the CD methods in the context of longitudinal GEE regression analysis, including two major scenarios. Scenario A is designed to evaluate (i) the validity of inference, (ii) scalability, (iii) efficiency, and (v) computational stability, where data are generated from a certain correlation structure. Scenario B is designed to demonstrate the robustness of the Rao-CD estimator against contaminated data or heterogeneous correlation structures; this is an important aspect (iv), regarding the advantage of the CD meta estimation. The estimation results from the Rao-CD method are compared in both scenarios with those obtained from the AEE method \cite[]{lin2011aggregated}, denoted by $\hat{\btheta}_{AEE}$, and full-data estimator obtained by processing the entire data once, $\hat{\btheta}_{full}$. 

{\bf Scenario A.} We consider a linear model $y_{k, ij} = \theta_0 + x_{k, ij}\theta_1 + \varepsilon_{k, ij}, i = 1, \dots, n_k, j = 1,\dots, l$, where $\btheta = (\theta_0, \theta_1)^{T} = (1/3, 1/2)^{T}$, $x_{k,ij} \sim N(0,1)$, and $\bvarepsilon_{k, i} = (\varepsilon_{k, i1}, \dots, \varepsilon_{k, il})^{T} \stackrel{i.i.d.}{\sim} \mathcal{N}\left(\bm{0}, \sigma^2R_{AR}(\rho)\right) $, with $R_{AR}(\rho)$ being an AR-1 correlation matrix whose $(i,j)^{th}$ element is $\rho^{|i-j|}$. The size of the entire data is fixed at $n = 100\hspace{0.1cm}000$, and the number of parallel datasets is set at $K = 5, 20, 50, 100, 200$, respectively, corresponding to the size of sub-dataset equal to $20\hspace{0.1cm}000, 5\hspace{0.05cm}000, 2\hspace{0.05cm}000, 1\hspace{0.05cm}000 \hspace{0.1cm} \text{and} \hspace{0.1cm} 500$.  
For the $k^{th}$ sub-dataset, estimating equation, $ n^{-1}_k\sum_{i=1}^{n_k}\bx_{k,i}^{T}R^{-1}_{AR}(\rho)\left(\by_{k, i} - \bx_{k, i}\btheta\right) = \bm{0},$ is used to estimate $\btheta$, where 
$\by_{k, i} = (y_{k, i1}, \dots, y_{k, il})^{T}$, and $\bx_{k, i} = (\bx_{k, i1}, \dots, \bx_{k, il})^{T}$, with $\bx_{k, ij} = (1, x_{k, ij})^{T}, j = 1, \dots, l$. The correlation parameter $\rho$ is consistently estimated by a method of moments suggested by \cite{liang1986longitudinal}.

Set $\rho \in \{0, 0.5, 0.8\}$. Note that since the true AR-1 correlation structure is used, the resulting GEE estimator $\hat{\btheta}_k$ is fully efficient for each sub-dataset. The results of summary statistics over 500 rounds of simulations are reported in Table~\ref{tab:lon_ho_gnn_re1}, including ABIAS, ASE, CP, ARE and ``PRE $<$ 1". Clearly, the CP by the Rao-CD method is close to the nominal $95\%$ level. This, together with the fact that ESE and ASE are comparable, provides numerical evidence for the validity of the asymptotic formulas in Theorems~\ref{the:cef_norm} and~\ref{the:cefk} as well as the one-step updating scheme given in Section~\ref{sec:imp}. With no surprise, the ARE of $\hat{\btheta}_{rcd}$ to $\hat{\btheta}_{full}$ or the ARE of $\hat{\btheta}_{AEE}$ to $\hat{\btheta}_{full}$ is around $1$ because the score function is used as the estimating function, confirming the result of efficiency equality shown in Theorem~\ref{the:corie}. However, both AREs decrease slightly as the size of sub-dataset, $m$, decreases. It is interesting to notice that although the ARE of $\hat{\btheta}_{rcd}$ to $\hat{\btheta}_{full}$ is close to 1, the ``PRE $<$ 1" is nearly $100\%$, instead of $50\%$, indicating that variance of $\hat{\btheta}_{rcd}$ is always smaller than that of $\hat{\btheta}_{full}$. This is because the correlation parameter $\rho$ is estimated in each sub-dataset, and such local estimate instead of a global estimate, appears to reduce variability in the CD method as pointed out in subsection~6.2.  In contrast, the ``PRE $<$ 1" of 
$\hat{\btheta}_{AEE}$ to $\hat{\btheta}_{full}$ has an opposite direction; it decreases as the magnitude of $\rho$ increases from 0 to 0.8. This indicates that using the sensitivity matrix in the weighting scheme for $\hat{\btheta}_{AEE}$ fails to pick up heterogeneous correlations effectively to lower down variability of the AEE estimation. 

\begin{figure}
	{\centering \includegraphics[width = 1\linewidth, height = 10cm]{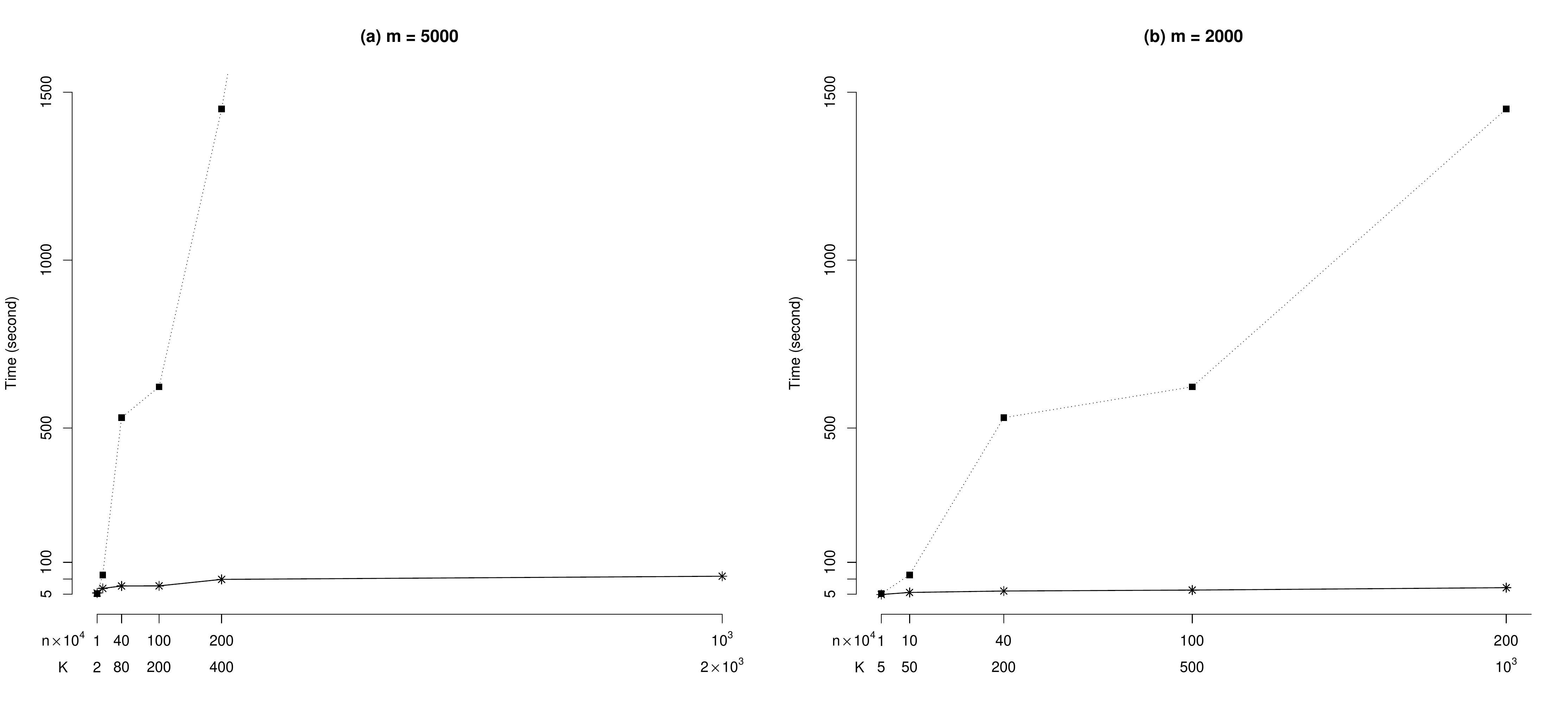}}
	\caption{The median computation time over 500 replicates for the proposed Rao-CD meta estimation (star) and full-data estimation (square) as $n$ increases. The size of each sub-dataset is fixed equally at $m = 2000$ and $5000$, respectively, while $K$ increases along with the increase in $n$. The full-data estimation $\hat{\btheta}_{full}$ fails to produce results when $n = 10^7$ due to computing memory limitations and the related results are not reported.}
	\label{fig:time}
\end{figure}

Figure~\ref{fig:time} displays a comparison of computation time between the Rao-CD $\hat{\btheta}_{rcd}$ and the full-data estimator $\hat{\btheta}_{full}$ as $n$ increases, while holding $m$ fixed at $m = 2\hspace{0.05cm}000$ and $5\hspace{0.05cm}000$. We see that the computational burden increases sharply for the full-data estimation $\hat{\btheta}_{full}$ as $n$ increases, whereas the computation time for the Rao-CD meta estimator $\hat{\btheta}_{rcd}$ remains almost unchanged and very low, which clearly demonstrates the scalability of the proposed CD estimation methods. Computation time for $\hat{\btheta}_{full}$ with $n = 10^7$ is not reported because the related computation exceeds the maximum memory limit allowed on the high performance Linux cluster used in our simulation study.

\begin{table}
	\hspace{-2cm}
	\caption{\label{tab:lon_ho_gnn_re1} The ABIAS, ESE, ASE and CP for the slope parameter of the GEE model with longitudinal normal data under Scenario A, including Rao-CD $\hat{\btheta}_{rcd}$, AEE $\hat{\btheta}_{AEE}$ and FULL $\hat{\btheta}_{full}$ under $K = 5, 20, 50, 100, 200$ with $n = 100, 000$ and $\rho = 0, 0.5, 0.8$. In addition, the ARE and the proportion of relative efficiency less than 1 (PRE $<$ 1) are listed.}
	
	\hspace{-2cm}
	\resizebox{\textwidth}{!}{
		\begin{tabular}{ccl*{14}{c}}
			\hline\hline
			$m$&$K$&&\multicolumn{3}{c}{$\rho=0$}&&\multicolumn{3}{c}{$\rho=0.5$}&&\multicolumn{3}{c}{$\rho = 0.8$}\\
			\hline
			&&&Rao-CD&AEE&FULL&&Rao-CD&AEE&FULL&&Rao-CD&AEE&FULL\\
			$20000$&5   &ABIAS$\times 10^{-3}$&1.099&1.100&1.100&&0.890&0.891&0.890&&0.561&0.562&0.562\\
			&&ESE$\times 10^{-3}$  &1.387&1.414&1.388&&1.125&1.125&1.125&&0.711&0.711&0.710\\
			&&ASE$\times 10^{-3}$  &1.414&1.414&1.414&&1.142&1.142&1.142&&0.721&0.721&0.721\\
			&&CP   &0.946&0.946&0.946&&0.936&0.936&0.938&&0.938&0.938&0.938\\
			&&ARE  &1.000&1.000&1.000&&1.000&1.000&1.000   &&1.000&1.000&1.000\\
			&&PRE $<$ 1 &53.8 &51.6 &--   &&99.8  &66.0 &--   &&100  &50.4 &--\\
			$5000$&20   &ABIAS$\times 10^{-3}$&1.097&1.100&1.100&&0.887&0.891&0.890&&0.560&0.562&0.562\\
			&&ESE$\times 10^{-3}$  &1.386&1.414&1.388&&1.123&1.124&1.125&&0.710&0.711&0.710\\
			&&ASE$\times 10^{-3}$  &1.413&1.414&1.414&&1.141&1.142&1.142&&0.721&0.721&0.721\\
			&&CP   &0.948&0.946&0.946&&0.938&0.936&0.938&&0.938&0.936&0.938\\
			&&ARE  &0.999&1.000&1.000   &&0.999&1.000&1.000   &&0.999&1.000&1.000\\ 
			&&PRE $<$ 1 &61.0 &54.6 &--   &&100  &77.8  &--   &&100  &47.4  &--\\
			$2000$&50   &ABIAS$\times 10^{-3}$&1.098&1.100&1.100&&0.889&0.892&0.890&&0.561&0.563&0.562\\
			&&ESE$\times 10^{-3}$  &1.387&1.413&1.388&&1.125&1.126&1.125&&0.711&0.712&0.710\\
			&&ASE$\times 10^{-3}$  &1.412&1.414&1.414&&1.140&1.142&1.142&&0.720&0.721&0.721\\
			&&CP   &0.944&0.946&0.946&&0.938&0.938&0.938&&0.934&0.940&0.938\\
			&&ARE  &0.999&1.000&1.000   &&0.999&1.000&1.000   &&0.999&1.000&1.000\\
			&&PRE $<$ 1 &76.4 &58.6 &--   &&100  &90.4  &--   &&100  &41.8  &--\\
			$1000$&100  &ABIAS$\times 10^{-3}$&1.099&1.100&1.100&&0.891&0.892&0.890&&0.563&0.563&0.562\\
			&&ESE$\times 10^{-3}$  &1.389&1.413&1.388&&1.128&1.125&1.125&&0.713&0.711&0.710\\
			&&ASE$\times 10^{-3}$  &1.410&1.414&1.414&&1.139&1.142&1.142&&0.719&0.721&0.721\\
			&&CP   &0.944&0.946&0.946&&0.936&0.938&0.938&&0.936&0.938&0.938\\
			&&ARE  &0.997&0.999&1.000   &&0.997&1.000&1.000   &&0.997&1.000&1.000\\
			&&PRE $<$ 1 &91.0 &67.2 &--   &&100  &97.4  &--   &&100  &37.6  &--\\ 
			$500$&200   &ABIAS$\times 10^{-3}$&1.101&1.100&1.100&&0.890&0.891&0.890&&0.563&0.562&0.562\\
			&&ESE$\times 10^{-3}$  &1.391&1.412&1.388&&1.132&1.125&1.125&&0.716&0.710&0.710\\
			&&ASE$\times 10^{-3}$  &1.406&1.414&1.414&&1.136&1.141&1.142&&0.717&0.721&0.721\\
			&&CP   &0.946&0.946&0.946&&0.936&0.938&0.938&&0.932&0.936&0.938\\
			&&ARE  &0.994&0.998&1.000   &&0.994&0.999&1.000   &&0.994&1.000&1.000\\
			&&PRE $<$ 1 &99.8 &79.2 &--   &&100  &99.8  &--   &&100  &35.2  &--\\ 
			\hline\hline
		\end{tabular}
	}
\end{table}

{\bf Scenario B.} To demonstrate the robustness of the Rao-CD estimation method, we first consider the case of data contamination. Under the same GEE model as given in Scenario A with a fixed $\rho = 0.5$, we generate contaminated data cases as follows: one outlier in one subject's response vector is introduced by using $100y_{k, ij}$, where $y_{k, ij}$ is a randomly selected data point from the vector of repeated measurements for subject $i$. The proportions of contaminated subjects are chosen to be $0.1\%$ (denoted as P1) and $0.2\%$ (denoted as P2). Two schemes of outlier allocations are considered: (i) Random allocation refers to the case where subjects with outliers appear randomly in any of $K$ sub-datasets; and (ii) fixed allocation corresponds to the case where subjects with outliers enter only one chosen sub-dataset. The size of entire data is set at $n = 2\hspace{0.05cm}000$ and $10\hspace{0.05cm}000$. The results of summary statistics over 500 replications are reported in Table~\ref{tab:out}, with the following highlights. 
\begin{itemize}
	\item ABIAS, ESE and ASE of $\hat{\btheta}_{full}$ or $\hat{\btheta}_{AEE}$ are 2 to 3 times larger than those of $\hat{\btheta}_{rcd}$; the latter estimator appears remarkably stable and robust.
	
	\item When the size of entire data is moderate like $n = 2\hspace{0.05cm}000$ (the bottom block of Table~\ref{tab:out}), light data contamination (P1) has minor effects on all the three estimation methods in terms of coverage probability. Coverage probabilities given by $\hat{\btheta}_{full}$ or $\hat{\btheta}_{AEE}$ are slightly over $95\%$ due to their wider confidence intervals caused by larger ASEs. In contrast, the coverage probabilities given by the proposed Rao-CD method appear reasonably close to $95\%$. 
	
	\item As the sample size increases, {\em say}, $n = 10\hspace{0.05cm}000$ (the top block of Table~\ref{tab:out}), where more outliers are present in the data, the coverage probabilities by $\hat{\btheta}_{rcd}$ remain robustly close to $95\%$; in contrast, the coverage probabilities by $\hat{\btheta}_{full}$ and $\hat{\btheta}_{AEE}$ decrease to $88\%$ and $70\%$ corresponding to the cases of $0.1\%$ and $0.2\%$ subjects with outliers, respectively. The main reason for such poor coverages is rooted in the fact that both $\hat{\btheta}_{full}$ and $\hat{\btheta}_{AEE}$ suffer severely nonignorable estimation biases caused by outliers. 
\end{itemize}
In summary, the above findings provide supporting evidence to the automatic downweighting strategy and data dilution phenomenon for the Rao-CD method. While the former has been reported in the literature \cite[{\emph{ e.g.}}][]{qu2004assessing}, the latter is uniquely related to data partition. All three methods are affected by an increased number of outliers, so theoretically understanding respective breakpoints for these methods is of great interest, which is one of future research directions. 

\begin{table}
	\hspace{-0.5cm}
	\caption{\label{tab:out} The ABIAS, ESE, ASE, CP, ARE and PRE $<$ 1 for the slope parameter of the GEE model with longitudinal normal data under Scenario B, where the proportions of subjects with contaminated data are $0.1\%$ (P1) and $0.2\%$ (P2), respectively. Two schemes of outlier allocations, random and fixed, are considered.}
	\hspace{-1.5cm}
	\resizebox{15cm}{!}{
		\begin{tabular}{cl*{9}{c}}
			\hline\hline
			&&\multicolumn{9}{c}{$n = 10, 000$}\\
			&&\multicolumn{4}{c}{$K = 50$}&&\multicolumn{4}{c}{$K = 20$}\\
			&&\multicolumn{2}{c}{Random}&\multicolumn{2}{c}{Fixed}&&\multicolumn{2}{c}{Random}&\multicolumn{2}{c}{Fixed}\\
			&&P1&P2&P1&P2&&P1&P2&P1&P2\\
			\hline
			FULL &ABIAS$\times 10^{-3}$&11.320 &19.650&11.018 &20.641&&11.320 &19.650&10.995 &19.527\\
			&ESE$\times 10^{-3}$  &9.498&12.400&8.929 &12.593&&9.498&12.400&8.649 &12.754\\
			&ASE$\times 10^{-3}$  &9.225&12.223&8.861&12.578&&9.225&12.223&8.989&12.161\\
			&CP     &0.896 &0.706 &0.892 &0.684 &&0.896 &0.706 &0.894 &0.716 \\
			&ARE    &1.000 &1.000 &1.000 &1.000 &&1.000 &1.000 &1.000 &1.000\\
			&PRE $<$ 1   &--    &--   &--   &--   &   &--   &--   &--   &--   \\
			
			AEE  &ABIAS$\times 10^{-3}$  &7.968&14.331&7.455&13.519&&8.534&15.974&7.446&13.015 \\
			&ESE$\times 10^{-3}$    &7.011&9.140&6.422&8.551&&7.416&10.192&6.178&8.797 \\
			&ASE$\times 10^{-3}$    &6.824&9.128&6.291&8.486&&7.231&10.072&6.446&8.373 \\
			&CP     &0.882 &0.720 &0.888 &0.698 &&0.884 &0.712 &0.898 &0.742       \\
			&ARE    &0.755 &0.752 &0.727 &0.680 &&0.798 &0.827 &0.734 &0.695\\
			&PRE $<$ 1   &100   &100     &100     &100     &&100     &100  &100   &100\\
			
			Rao-CD&ABIAS$\times 10^{-3}$&3.111 &3.252 &2.992 &2.995 &&3.317 &3.620 &3.029 &3.030\\  
			&ESE$\times 10^{-3}$ &3.935 &4.092 &3.714 &3.718 &&4.124 &4.486 &3.775 &3.774  \\
			&ASE$\times 10^{-3}$ &3.750 &3.948 &3.595 &3.596 &&4.028 &4.502 &3.681 &3.683  \\
			&CP     &0.944 &0.938 &0.940 &0.942 &&0.940 &0.936 &0.952 &0.954   \\
			&ARE    &0.460 &0.351 &0.455 &0.310 &&0.492 &0.399 &0.459 &0.332\\
			&PRE $<$ 1   &100   &100   &100   &100   &&100   &100   &100   &100\\
			
			\hline
			&&\multicolumn{9}{c}{$n = 2, 000$}\\
			&&\multicolumn{4}{c}{$K = 10$}&&\multicolumn{4}{c}{$K = 4$}\\
			FULL &ABIAS$\times 10^{-3}$   &15.761&24.831&15.035&27.249&&15.761&24.831&16.561&25.401\\
			&ESE$\times 10^{-3}$
			&19.773&27.187&19.428&28.690&&19.773&27.187&20.375&27.345\\
			&ASE$\times 10^{-3}$
			&17.831&24.743&17.577&25.803&&17.831&24.743&18.095&25.321\\
			&CP     &0.968 &0.970 &0.968 &0.972 &&0.968 &0.970 &0.954 &0.964\\
			&ARE    &1.000 &1.000 &1.000 &1.000 &&1.000 &1.000 &1.000 &1.000\\
			&PRE $<$ 1   &--    &--   &--   &--   &   &--   &--   &--   &--   \\
			AEE  &ABIAS$\times 10^{-3}$ &11.873&18.900&11.268&18.761&&12.766&21.160&12.536&18.563\\
			&ESE$\times 10^{-3}$
			&14.723&20.901&14.276&20.224&&15.665&23.586&15.274&19.742\\
			&ASE$\times 10^{-3}$
			&13.803&18.987&13.386&18.383&&14.740&21.159&14.230&18.964\\
			&CP     &0.966 &0.956 &0.962 &0.956 &&0.972 &0.964 &0.962 &0.960\\
			&ARE    &0.819 &0.792 &0.815 &0.737 &&0.864 &0.869 &0.830 &0.773\\
			&PRE $<$ 1   & 94.8 &99    &95.6  &99.8  &&94.2  &96.6  &95.8  &100\\
			
			Rao-CD &ABIAS$\times 10^{-3}$ &6.843&7.150&6.619&6.749&&7.260&8.081&7.056&7.254\\
			&ESE$\times 10^{-3}$
			&8.618&9.017&8.416&8.541&&9.190&10.435&8.990&9.092\\
			&ASE$\times 10^{-3}$
			&8.418&8.866&8.282&8.363&&9.172&10.391&8.860&9.088\\
			&CP     &0.942 &0.950 &0.940 &0.944 &&0.952 &0.952 &0.936 &0.940\\
			&ARE    &0.609 &0.466 &0.616 &0.421 &&0.653 &0.532 &0.629 &0.469\\
			&PRE $<$ 1   &99.2   &100   &100   &100   &&98.2 &99.8  &98.4   &100\\
			
			\hline\hline
		\end{tabular}
	}
\end{table}

Now we turn to the evaluation of robustness against heterogeneous correlation structures. The following insight is critical to understand the performance of these methods. That is, according to \cite{liang1986longitudinal}, the GEE estimator remains consistent even if the correlation matrix is misspecified. We create the full dataset by merging $Q$ datasets generated from the GEE linear models, each with a correlation structure randomly selected from independence, AR-1 or compound symmetry (CS) with correlation parameter $\rho_q \sim U(0.1, 0.9)$ for $q = 1, \dots, Q$. Then, the integrated dataset is randomly partitioned into $K$ sub-datasets. Set $Q = 5, 50$ and $100$. In the GEE analysis, we always use AR-1 working correlation. Table~\ref{tab:lon_he_gnn_re1n} reports the simulation results. Additional results obtained under CS or independence working correlation are not reported here due to the fact that they are very similar to those shown in Table~\ref{tab:lon_he_gnn_re1n}. From Table~\ref{tab:lon_he_gnn_re1n}, we see that all the three estimation methods have shown proper coverage probabilities, being close to the nominal level $95\%$. The Rao-CD method appears more efficient than the other two methods, $\hat{\btheta}_{AEE}$ and $\hat{\btheta}_{full}$, with ARE of $\hat{\btheta}_{rcd}$ to $\hat{\btheta}_{full}$ equal to 92.7\% even when the size of sub-dataset is as large as $m = 20\hspace{0.05cm}000$. This is in agreement with the theoretical results established in Theorem~\ref{the:corie}: $\hat{\btheta}_{rcd}$ is more efficient than $\hat{\btheta}_{full}$ when the estimation function is not the score function where the homogeneous asymptotic Godambe information assumption does not hold.

\begin{table}
	\hspace{-2cm}
	\caption{	\label{tab:lon_he_gnn_re1n}
		The ABIAS, ESE, ASE, CP, ARE and PRE $<$ 1 for the slope parameter of the GEE model with longitudinal normal data under Scenario B, where Rao-CD $\hat{\btheta}_{rcd}$, AEE $\hat{\btheta}_{AEE}$ and FULL $\hat{\btheta}_{full}$ are compared in the setting of $K = 5, 20, 50, 100, 200$ with $n = 100, 000$. The full data is created by merging  $Q = 5, 50, 100$ sub-datasets with different correlation structures.}
	\hspace{-2cm}
	\resizebox{\textwidth}{!}{
		\begin{tabular}{ccl*{12}{c}}
			\hline\hline
			$m$&$K$&&\multicolumn{3}{c}{$Q = 5$}&&\multicolumn{3}{c}{$Q = 50$}&&\multicolumn{3}{c}{$Q = 100$}\\
			&&&Rao-CD&AEE&FULL&&Rao-CD&AEE&FULL&&Rao-CD&AEE&FULL\\
			\hline
			$20000$&5&ABIAS$\times 10^{-3}$
		&1.015&1.037&1.061&&1.025&1.038&1.014&&1.014&1.024&1.021\\
			         &&  ESE$\times 10^{-3}$
	    &1.268&1.309&1.319&&1.279&1.292&1.270&&1.292&1.302&1.299\\
			         &&  ASE$\times 10^{-3}$
	    &1.211&1.253&1.305&&1.310&1.318&1.321&&1.315&1.320&1.321\\
			         &&  CP
	    &0.946&0.938&0.958&&0.956&0.960&0.960&&0.948&0.940&0.944\\
			         && ARE 
		&0.927&0.959&1.000&&0.992&0.998&1.000&&0.996&0.999&1.000\\
			         &&   PRE $<$ 1&  99.4 & 71.2 &--   &&  97.6&  59.4&-- &&  99.0&  59.6& --\\
			$5000$&20 &ABIAS$\times 10^{-3}$
		&1.001&1.054&1.061&&0.989&1.038&1.014&&1.011&1.048&1.021\\
			         &&  ESE$\times 10^{-3}$
	    &1.267&1.331&1.319&&1.240&1.299&1.270&&1.279&1.317&1.299\\
			         &&  ASE$\times 10^{-3}$
		&1.166&1.228&1.305&&1.265&1.301&1.321&&1.293&1.313&1.321 \\
			         &&  CP
	    &0.934&0.934&0.958&&0.948&0.950&0.960&&0.948&0.940&0.944\\
			         &&  ARE
	    &0.892&0.940&1.000&&0.958&0.985&1.000&&0.979&0.994&1.000\\
			         &&  PRE $<$ 1&  99.8 & 72.6 &--   &&  100&  80.2&-- &&  100&  74.6& --\\
			$2000$&50 &ABIAS$\times 10^{-3}$
		&0.982&1.026&1.061&&0.920&1.010&1.014&&0.986&1.037&1.021\\
			         &&  ESE$\times 10^{-3}$
	    &1.240&1.294&1.319&&1.153&1.256&1.270&&1.249&1.298&1.299\\
			         &&  ASE$\times 10^{-3}$
	    &1.153&1.218&1.305&&1.193&1.278&1.321&&1.249&1.296&1.321\\
			         &&  CP
	    &0.926&0.936&0.958&&0.956&0.966&0.960&&0.942&0.944&0.944\\
			         && ARE 
	    &0.882&0.932&1.000&&0.903&0.968&1.000&&0.945&0.981&1.000\\
			         &&   PRE $<$ 1&  99.8 & 73.2 &--   &&  100&  89.6&-- &&  100&  88.4& --\\
			$1000$&100&ABIAS$\times 10^{-3}$
		&0.984&1.028&1.061&&0.889&1.022&1.014&&0.925&0.978&1.021 \\
			         &&  ESE$\times 10^{-3}$
	    &1.241&1.293&1.319&&1.134&1.270&1.270&&1.162&1.222&1.299 \\
			         &&  ASE$\times 10^{-3}$
	    &1.149&1.216&1.305&&1.147&1.263&1.321&&1.190&1.279&1.321\\
			         &&  CP
	    &0.920&0.934&0.958&&0.948&0.956&0.960&&0.934&0.948&0.944\\
			         &&  ARE
	    &0.879&0.930&1.000&&0.868&0.956&1.000&&0.901&0.968&1.000\\
			         &&   PRE $<$ 1&  99.8 & 73.8 &--   &&  100&  95.0&-- &&  100&  94.6& --\\
			$500$&200 &ABIAS$\times 10^{-3}$
		&0.996&1.034&1.061&&0.872&1.009&1.014&&0.903&0.993&1.021\\
			         &&  ESE$\times 10^{-3}$
	    &1.253&1.301&1.319&&1.112&1.260&1.270&&1.134&1.248&1.299\\
			         &&  ASE$\times 10^{-3}$
		&1.142&1.212&1.305&&1.123&1.252&1.321&&1.140&1.262&1.321\\
			         &&  CP
	    &0.928&0.934&0.958&&0.952&0.950&0.960&&0.944&0.958&0.944\\
			         && ARE
	    &0.873&0.928&1.000&&0.850&0.948&1.000&&0.863&0.955&1.000\\			
				     &&   PRE $<$ 1&  100 & 74.2 &--   &&  100&  96.8&-- &&  100&  98.4& --\\
						         
			\hline\hline
		\end{tabular}
	}
\end{table}

\subsection{Cox regression model}
In this subsection, we focus only on the evaluation of the robustness against heterogeneous baseline hazards. We consider the following Cox Proportional Hazards model,
$\lambda(t) = \lambda_0(t)\exp(x_1\theta_1 + x_2\theta_2)$, where $\btheta = (\theta_1, \theta_2)^{T} = (1/3, 1/2)^{T}$ is the same across K sub-datasets and 
$\lambda_0(t) = \lambda\rho t^{\rho-1}$ is the hazard of a Weibull distribution with parameters $\lambda$ and $\rho$. The censoring indicator $1 - \delta$ is generated from Binomial(0.3), which indicates the censoring probability is around $30\%$. Two situations of baseline hazard heterogeneities are considered, 
\begin{itemize}
	\item[H1.] One simulated full dataset is combined from 2 sub-datasets, each with respectively fixed parameters that are randomly generated from $\lambda_q \sim U[0.5, 5]$ and 
	$\rho_q \sim U[0.5, 5], q = 1, 2$. The data are then divided into $K$ sub-datasets in light of the designed group structure: when $K$ is an even number, there is no sub-dataset containing data from two groups; when $K$ is an odd number, there is one sub-dataset with data from two groups. 
	
	\item[H2.] One simulated full dataset is generated with the baseline hazard with fixed parameters $(\lambda, \rho) = (1, 1)$. Then, $25\%$ of the data are randomly selected and replaced by those generated from a similar setting with randomly generated parameters that are randomly generated from $\lambda \sim U[0.5, 5]$ and $\rho \sim U[0.5, 5]$. Then the data are randomly divided into $K$ sub-datasets.
\end{itemize}
 
The size of the full data is chosen to be $n = 4\hspace{0.05cm}000$ and $n = 10\hspace{0.05cm}000$ with $m = 500, 1\hspace{0.05cm}000, 2\hspace{0.05cm}000$. The analysis results for $\theta_1$ are reported in Table~\ref{tab:sur_re1}. Additional results for $\theta_2$ are similar, and thus not shown here. For case H2, all the three methods, $\hat{\btheta}_{rcd}$, $\hat{\btheta}_{AEE}$ and $\hat{\btheta}_{full}$, have yielded coverage probabilities around the nominal level $95\%$. ARE of $\hat{\btheta}_{rcd}$ to $\hat{\btheta}_{full}$ or ARE of $\hat{\btheta}_{AEE}$ to $\hat{\btheta}_{full}$ is slightly bigger than 1. These results are aligned with our discussion in subsection~6.3 that more variabilities are resulted from a larger parameter space of the baselines hazards $\lambda_{0, k}(t)$, which is not explicitly used in the estimation. For case H1, $\hat{\btheta}_{full}$ is biased, resulting in a significantly low coverage probability, only at $35\%$ level. $\hat{\btheta}_{rcd}$ or $\hat{\btheta}_{AEE}$ has produced reliable coverage probabilities with $K$ being an even number except the case of $K = 5$ an odd number. When $K = 5$ as pointed out above, there exists one sub-dataset, half of which is from one group and the other half is from the other group. In this case, the common baseline hazard is violated. Thus, the resulted estimates are biased, leading to a poor coverage. These numerical results give rise to a mixed message: in some cases, the Rao-CD method is robust, whereas in other case it is not robust to baseline heterogeneity. More systematic investigation on this issue is required in future research. 

\begin{table}
	\hspace{-2cm}
	\caption{	\label{tab:sur_re1}
		The ABIAS, ESE, ASE, CP, ARE and PRE $<$ 1 for $\theta_1$ of the Cox Proportional Hazard model under two situations of heterogeneous baseline hazards, denoted by H1 and H2, where Rao-CD $\hat{\btheta}_{rcd}$, AEE $\hat{\btheta}_{AEE}$ and FULL $\hat{\btheta}_{full}$ are compared. The full data size is fixed at $n = 2000$ and $n = 10, 000$ with $m = 500, 1000, 2000$. }\\
	\hspace{-2cm}
	\resizebox{13cm}{!}{
		\begin{tabular}{ccl*{8}{c}}
			\hline\hline
			&&&\multicolumn{2}{c}{$m = 500$}&&\multicolumn{2}{c}{$m=1000$}&&\multicolumn{2}{c}{$m = 2000$}\\
			&&&$K = 8$&$K = 20$&&$K = 4$&$K = 10$&&$K = 2$&$K = 5$\\
			\hline
			&Rao-CD&ABIAS&0.016&0.010&&0.016&0.010&&0.016&0.018\\
			&&ESE  &0.020&0.013&&0.020&0.013&&0.020&0.017\\
			&&ASE  &0.021&0.013&&0.021&0.013&&0.021&0.013\\
			&&CP   &0.948&0.958&&0.944&0.962&&0.946&0.732\\
			&&ARE  &1.028&1.032&&1.027&1.031&&1.027&1.024\\
			&&PRE $<$ 1 &  6.8&2.2  && 7.0&2.2  && 7.4&3.4\\
			&AEE&ABIAS&0.016&0.010&&0.016&0.010&&0.016&0.018\\
			&&ESE  &0.020&0.013&&0.020&0.013&&0.020&0.017\\
			H1 &&ASE  &0.021&0.013&&0.021&0.013&&0.021&0.013\\
			&&CP   &0.952&0.962&&0.946&0.966&&0.946&0.736\\
			&&ARE  &1.032&1.036&&1.029&1.033&&1.027&1.025\\
			&&PRE $<$ 1 &  3.4&  0.4&& 4.6&  1.6&& 5.6&2.2\\
			&FULL&ABIAS&0.065&0.070&&0.065&0.070&&0.065&0.070\\
			&&ESE  &0.049&0.049&&0.049&0.049&&0.049&0.049\\
			&&ASE  &0.020&0.013&&0.020&0.013&&0.020&0.013\\
			&&CP   &0.356&0.214&&0.356&0.214&&0.356&0.214\\
			&&ARE  &1.000&1.000&&1.000&1.000&&1.000&1.000\\
			&&PRE $<$ 1 & --   &--   &&    --&   -- &&   --&  --\\
			\hline
			&Rao-CD&ABIAS&0.016&0.010&&0.016&0.010&&0.016&0.010\\
			&&ESE  &0.021&0.013&&0.020&0.013&&0.020&0.013\\
			&&ASE  &0.021&0.013&&0.021&0.013&&0.021&0.013\\
			&&CP   &0.956&0.952&&0.950&0.954&&0.952&0.962\\
			&&ARE  &1.001&1.002&&1.001&1.001&&1.000&1.001\\
			&& PRE $<$ 1& 39.4& 31.4&& 40.2& 32.0&& 41.2&35.4\\
			&AEE&ABIAS&0.016&0.010&&0.016&0.010&&0.016&0.010\\
			&&ESE  &0.020&0.013&&0.020&0.013&&0.020&0.013\\
			H2 &&ASE  &0.021&0.013&&0.021&0.013&&0.021&0.013\\
			&&CP   &0.956&0.962&&0.954&0.954&&0.954&0.960\\
			&&ARE  &1.005&1.006&&1.003&1.003&&1.001&1.001\\
			&& PRE $<$ 1&  8.0&  2.6&& 20.2&  8.8&& 29.0&15.8\\
			&FULL&ABIAS&0.016&0.010&&0.016&0.010&&0.016&0.010\\
			&&ESE  &0.020&0.013&&0.020&0.013&&0.020&0.013\\
			&&ASE  &0.021&0.013&&0.021&0.013&&0.021&0.013\\
			&&CP   &0.952&0.956&&0.952&0.956&&0.952&0.956\\ 
			&&ARE  &1.000&1.000&&1.000&1.000&&1.000&1.000\\
			&&PRE $<$ 1 & --   &--    &&    --&   -- &&    --&  --\\
			\hline\hline
		\end{tabular}
	}
\end{table}

\section{Data examples}
\label{sec:real}
In this section, we present two real data analysis examples to illustrate the proposed Rao-CD meta estimation method. All numerical calculations are implemented by the Python and R software in the University of Michigan Hadoop platform, and the software packages are available for download at Song Lab webpage http://www.umich.edu/$\sim$songlab/software.html\#RCD.

\subsection{Quantile analysis of BMI data}
The Body Mass Index (BMI) data is from one of our collaborative projects, the Early Life Exposure in Mexico to Environmental Toxicants (ELEMENT), conducted in Mexico city. A total of $n = 1222$ children of ages 3-4 from four ELEMENT cohorts is included in this data analysis. In this application, the data is not randomly divided rather it is split according to the cohort formation. The central question of interest related to nutritional science is to evaluate how five dietary patterns (DP), $x_j, j = 1, \dots, 5$,  may be associated with chilren's quantiles of BMI. Here, $x_1$ represents vegetables and lean proteins, $x_2$ represents Maize products and sugar-sweetened beverages, $x_3$ represents processed meats and refined grains, $x_4$ represents fruit and yogurt, and $x_5$ represents whole grain and fat. In this association analysis, we adjust several important confounders, including maternal age ($x_6$), children's age ($x_7$, binary with 0 representing 3 years and 1 for 4 years), maternal education ($x_8$), children's gender ($x_9$, 0 for girl) and the number of kids in the family (parity, $x_{10}$). We consider quantile regression models with $\tau = 0.25 (Q1), 0.5 (Q2)$, and $0.75 (Q3)$, respectively, to understand the effects of dietary patterns on different BMI profiles of these Mexican children. Table~\ref{tab:ana1} lists estimated coefficients, asymptotic standard errors(ASEs), and $p$-values obtained by the Rao-CD, Wald-CD and full-data methods. When the CD method is used, the data are divided into 4 sub-datasets by their cohorts that are recruited at four different times, PL, BI, C1, and SF. The minimum size of those 4 sub-datasets is $m = 210$. Table~\ref{tab:ana1} shows that the Rao-CD method yields a highly consistent inference with the full-data method, concerning the association of BMI quantiles with the dietary patterns. However, the Wald-CD method exhibits some different inference results due possibly to its numerical instability. For example, at the $\tau = 0.25$ BMI quantile level, the Wald-CD method indicates that a higher intake of dietary pattern 5 (i.e. whole grain and fat) tends to have lower Q1 BMI quantile while both Rao-CD and full methods do not detect such significant association. At the median BMI, both Rao-CD and full methods suggest that higher intake of dietary pattern 5 associated with lower median BMI, while the Wald-CD method fails to capture this association. Similar numerical differences have been also seen from the simulation results reported in Table~\ref{tab:qu_robust}. 
  
\begin{table}
	\hspace{-1.5cm}
	\caption{	\label{tab:ana1}
		The estimates, ASEs and $p$-values for the coefficients in three quantile regression models for the BMI data with $\tau = 0.25, 0.5$ and $0.75$, respectively, where Rao-CD $\hat{\btheta}_{rcd}$, Wald-CD $\hat{\btheta}_{wcd}$ and full $\hat{\btheta}_{full}$ are listed. The full data is divided according to 4 cohorts, BI, C1, SF, and PL, when Rao-CD and Wald-CD methods are used.}
	\hspace{-1.5cm}
	\resizebox{18cm}{!}{
		\begin{tabular}{ll*{11}{r}}
			\hline\hline
			&&\multicolumn{3}{c}{$\tau = 0.25$}&&\multicolumn{3}{c}{$\tau = 0.5$}
			&&\multicolumn{3}{c}{$\tau = 0.75$}\\
			Covariates&&Rao-CD&Wald-CD&Full&&Rao-CD&Wald-CD&Full&&Rao-CD&Wald-CD&Full\\
			\hline
			DP1&EST&$-$0.109&$-$0.080&$-$0.129&&$-$0.113&$-$0.073&$-$0.091&&   0.025&   0.061&$-$0.007 \\
			&ASE&   0.033&   0.033&   0.045&&   0.040&   0.040&   0.044&&   0.049&   0.049&   0.052 \\
			&$p$-val&   0.001&   0.015&   0.004&&   0.005&   0.072&   0.040&&   0.602&   0.212&   0.887\\
			DP2&EST&$-$0.002&$-$0.053&$-$0.016&&$-$0.025&$-$0.079&$-$0.011&&$-$0.105&$-$0.080&$-$0.096 \\      
			&ASE&   0.028&   0.028&   0.045&&   0.042&   0.042&   0.047&&   0.026&   0.026&   0.051  \\
			&$p$-val&   0.947&   0.059&   0.729&&   0.548&   0.057&   0.811&&   0.000&   0.002&   0.057\\
			DP3&EST&$-$0.044&$-$0.132&$-$0.048&&$-$0.121&$-$0.124&$-$0.129&&$-$0.133&$-$0.058&$-$0.120 \\
			&ASE&   0.028&   0.028&   0.044&&   0.041&   0.041&   0.043&&   0.041&   0.041&   0.052       \\
			&$p$-val&   0.123&   0.000&   0.285&&   0.003&   0.002&   0.003&&   0.001&   0.154&   0.022\\
			DP4&EST&$-$0.136&$-$0.163&$-$0.148&&$-$0.088&$-$0.054&$-$0.098&&$-$0.190&$-$0.048&$-$0.190 \\
			&ASE&   0.030&   0.030&   0.041&&   0.042&   0.042&   0.041&&   0.041&   0.041&   0.050 \\
			&$p$-val&   0.000&   0.000&   0.000&&   0.038&   0.202&   0.016&&   0.000&   0.236&   0.000\\
			DP5&EST&$-$0.040&$-$0.061&$-$0.049&&$-$0.103&$-$0.057&$-$0.082&&$-$0.044&$-$0.009&$-$0.030\\
			&ASE&   0.028&   0.028&   0.044&&   0.046&   0.046&   0.041&&   0.043&   0.043&   0.052\\
			&$p$-val&   0.154&   0.029&   0.266&&   0.026&   0.223&   0.044&&   0.312&   0.841&   0.563\\
			\\
			AGE\_M&EST&   0.003&$-$0.001&$-$0.001&&   0.002&   0.001&   0.003&&   0.019&   0.003&   0.014 \\
			&ASE&   0.007&   0.007&   0.010&&   0.009&   0.009&   0.009&&   0.009&   0.009&   0.012  \\
			&$p$-val&   0.621&   0.919&   0.897&&   0.842&   0.925&   0.708&&   0.040&   0.720&   0.214 \\
			AGE\_C&EST&$-$0.244&$-$0.210&$-$0.278&&$-$0.023&$-$0.062&$-$0.073&&   0.356&   0.295&   0.251 \\
			&ASE&   0.087&   0.087&   0.169&&   0.172&   0.172&   0.166&&   0.153&   0.153&   0.189 \\
			&$p$-val&   0.005&   0.015&   0.099&&   0.895&   0.717&   0.661&&   0.020&   0.054&   0.185\\
			EDU   &EST&$-$0.015&   0.015&$-$0.001&&   0.018&   0.019&   0.016&&$-$0.008&   0.041&   0.018 \\
			&ASE&   0.014&   0.014&   0.016&&   0.015&   0.015&   0.015&&   0.017&   0.017&   0.020       \\
			&$p$-val&   0.260&   0.257&   0.944&&   0.234&   0.198&   0.280&&   0.621&   0.016&   0.356\\
			GENDER&EST&   0.088&   0.113&   0.092&&   0.111&   0.133&   0.060&&   0.184&   0.163&   0.088 \\
			&ASE&   0.069&   0.069&   0.086&&   0.083&   0.083&   0.091&&   0.084&   0.084&   0.106 \\
			&$p$-val&   0.198&   0.100&   0.285&&   0.182&   0.109&   0.509&&   0.029&   0.054&   0.403   \\
			PARITY &EST&$-$0.036&   0.041&$-$0.020&&$-$0.040&   0.046&$-$0.034&&$-$0.054&   0.045&$-$0.109 \\
			&ASE&   0.032&   0.032&   0.046&&   0.039&   0.039&   0.040&&   0.053&   0.053&   0.067  \\
			&$p$-val&   0.260&   0.198&   0.672&&   0.306&   0.246&   0.404&&   0.307&   0.399&   0.101  \\
			\hline\hline
		\end{tabular}
	}
\end{table}

\subsection{GEE analysis of clustered FARS data}
\begin{figure}
	{\centering \includegraphics[width = 1\linewidth, height = 20cm]{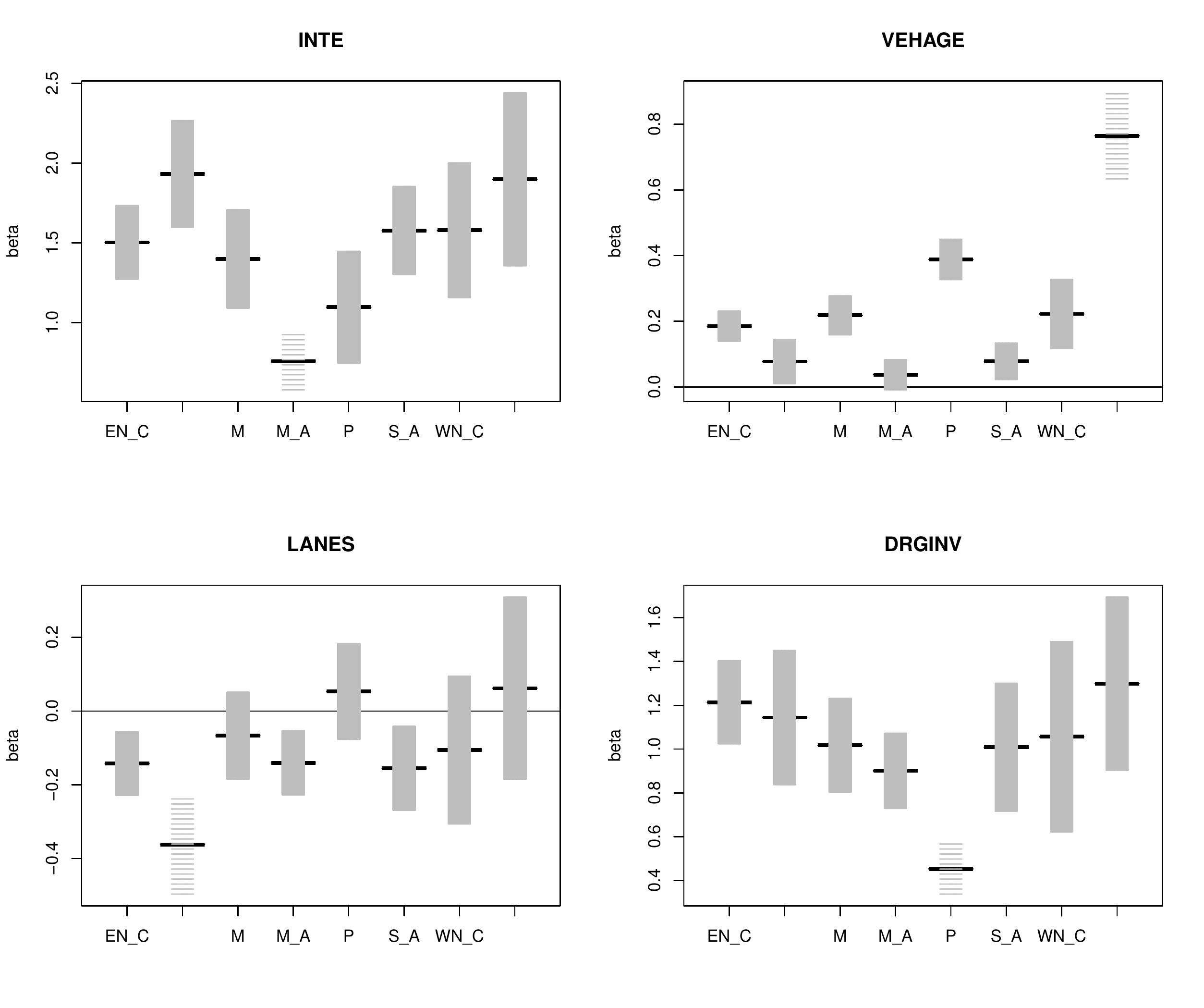}}
	\caption{The estimates and their $95\%$ confidence intervals for the clustered FARS data obtained by the GEE method from 8 geographic regions, East North Central (EN\_C), East South Central (ES\_C), Mountain (M), 
		Middle Atlantic (M\_A), Pacific (P), South Atlantic (S\_A), West North Central (WN\_C), and West South Central (WS\_C). }
	\label{fig:cds_division}
\end{figure}

\begin{figure}
	{\centering \includegraphics[width = 1\linewidth, height = 20cm]{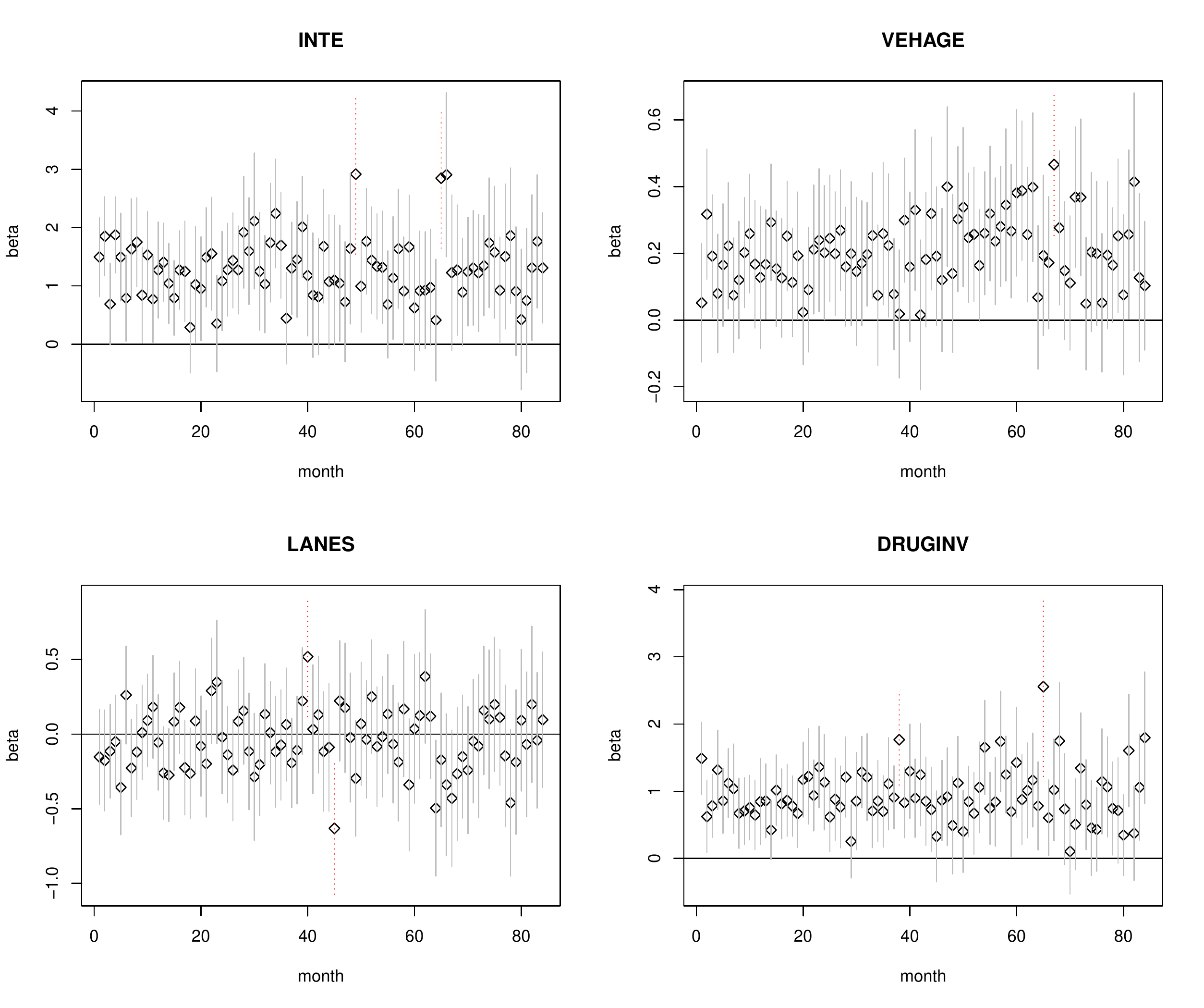}}
	\caption{The estimates and their $95\%$ confidence intervals for the clustered FARS data obtained by the GEE method from 84 individual sub-datasets, each for one month, during January, 2009 to December, 2015. }
	\label{fig:cds_month}
\end{figure}

Identifying risk factors associated with injury for subjects involving vehicle accidents is of great interest to policy makers and insurance companies. To address this question, we use a publicly available dataset from the National Highway and National Automotive Sampling System (NASS) Crashworthiness Data System (CDS) between January, 2009 and December 2015 across 8 regions (East North Central, East South Central, Moutain, Middle Atlantic, Pacific, South Atlantic, West North Central, West South Central) in the U.S. The dataset contains 54,794 occupants involving in 36,806 crash vehicles. In the data analysis, each vehicle is treated as a cluster, because occupants in a vehicle are more likely to be correlated for the chance of injury than those in other vehicles when an accident occurs. The cluster size varies from 1 to 8 with an average of 2 occupants. The response variable of interest is a binary variable of injury severity, 1 for a moderate or severe injury, and 0 for minor or no injury. We invoke the GEE logistic regression model with the compound symmetry  correlation structure to account for the within vehicle correlation. Fifteen potential risk factors are considered, including occupant's age, weight, height, speed limit (SPLIMIT), vehicle weight (VEGWGT), vehicle curb weight (CURBWGT), vehicle age (VEHAGE), air bag system deployed (BAGDEPLY, 1 for yes and 0 for no), police reported restraint use (PARUSE, 1 for used and 0 for no), occupant race (OCCRACE, 0 for white or black and 1 for else), number of lanes (LANES, 0 for $\leq 2$ and 1 for else), drug involvement in this accident (DRGINV, 1 for yes and 0 for no), driver's distraction/inattention to driving (DRIVDIST, 1 for attentive and 0 for else), roadway surface condition (SURCOND, 1 for dry and 0 for else), and had vehicle been in previous accidents (PREVACC, 0 for no and 1 for else). We divide the full data by 8 geographic regions or 84 months, which allows us to examine potential spatial or temporal differences in risk profiles of vehicle crashes. The minimum sizes of the resulting sub-datasets are 2321 and 372, respectively.
Table~\ref{tab:ana2} includes the estimated coefficients, standard errors and $p$-values obtained by the Rao-CD, AEE and full-data methods. Both Rao-CD and AEE methods yield highly consistent inference about these risk factors as the full-data method. There is only one difference; that is, the Rao-CD method suggests the number of lanes is associated with injury, but the other two methods disagree. 
To understand potential spartial or temporal heterogeneity, we also display four estimated coefficients and their standard errors across sub-datasets in Figures~\ref{fig:cds_division} and~\ref{fig:cds_month}. The individual estimates marked with zebra lines appear different from those obtained from other sub-datasets. For example, in Figure~\ref{fig:cds_division}, the estimation result in East South Central suggests that the number of lanes is negatively associated with the probability of moderate or severe injury, whereas the significant protection effect by this fact is not found in the other regions. Also, Figure~\ref{fig:cds_month} shows that there exist several outlying estimates from the set of estimates for one risk factor. Thus, the homogeneity assumption of regression coefficients needs to be relaxed, and it is worth a further investigation on how to integrate data properly so to accommodate potential data heterogeneity.     
 
\begin{table}
	\hspace{-1cm}
	\caption{	\label{tab:ana2}
		The estimates, ASEs, $p$-values for the regression coefficients of the GEE logistic regression model for the clustered FARS data, where Rao-CD $\hat{\btheta}_{rcd}$, AEE $\hat{\btheta}_{AEE}$ and full $\hat{\btheta}_{full}$ are listed. The full data is divided according to 8 geographic regions and 84 months, respectively, when Rao-CD or AEE method is used.}\\
	\hspace{0cm}
	\resizebox{14cm}{!}{
		\begin{tabular}{l*{12}{r}}
			\hline\hline
			&\multicolumn{11}{c}{Region}\\
			\cline{2-12}
			&\multicolumn{3}{c}{Rao-CD}&&\multicolumn{3}{c}{AEE}&&\multicolumn{3}{c}{Full}\\
			&EST&ASE&$p$-val&&EST&ASE&$p$-val&&EST&ASE&$p$-val\\
			\hline
			AGE & 0.173 & 0.011 & 0.000 && 0.174 & 0.011 & 0.000 && 0.178 & 0.010 & 0.000 \\ 
			WEIGHT & 0.154 & 0.012 & 0.000 && 0.159 & 0.013 & 0.000 && 0.160 & 0.012 & 0.000 \\ 
			HEIGHT & $-$0.038 & 0.012 & 0.002 && $-$0.043 & 0.012 & 0.000 && $-$0.042 & 0.012 & 0.000 \\ 
			SPLIMIT & 0.125 & 0.011 & 0.000 && 0.125 & 0.011 & 0.000 && 0.127 & 0.010 & 0.000 \\ 
			VEHWGT & 0.024 & 0.011 & 0.025 && 0.023 & 0.011 & 0.033 && 0.022 & 0.011 & 0.035 \\ 
			CURBWGT & $-$0.118 & 0.012 & 0.000 && $-$0.118 & 0.012 & 0.000 && $-$0.119 & 0.012 & 0.000 \\ 
			VEHAGE & 0.178 & 0.011 & 0.000 && 0.181 & 0.011 & 0.000 && 0.196 & 0.011 & 0.000 \\ 
			BAGDEPLY & 0.587 & 0.020 & 0.000 && 0.592 & 0.020 & 0.000 && 0.597 & 0.020 & 0.000 \\ 
			PARUSE & $-$1.098 & 0.039 & 0.000 && $-$1.098 & 0.040 & 0.000 && $-$1.134 & 0.038 & 0.000 \\ 
			OCCRACE & $-$0.011 & 0.022 & 0.603 && $-$0.014 & 0.022 & 0.520 && $-$0.016 & 0.021 & 0.465 \\ 
			LANES & $-$0.041 & 0.021 & 0.048 && $-$0.039 & 0.021 & 0.058 && $-$0.036 & 0.021 & 0.083 \\ 
			DRGINV & 0.883 & 0.034 & 0.000 && 0.877 & 0.034 & 0.000 && 0.884 & 0.033 & 0.000 \\ 
			DRIVDIST & $-$0.212 & 0.021 & 0.000 && $-$0.210 & 0.021 & 0.000 && $-$0.213 & 0.021 & 0.000 \\ 
			SURCOND & 0.136 & 0.025 & 0.000 && 0.138 & 0.025 & 0.000 && 0.145 & 0.025 & 0.000 \\ 
			PREVACC & $-$0.006 & 0.028 & 0.839 && $-$0.004 & 0.029 & 0.901 && $-$0.001 & 0.028 & 0.980 \\ 
\hline
	&\multicolumn{11}{c}{Month}\\
	\cline{2-12}
		&\multicolumn{3}{c}{Rao-CD}&&\multicolumn{3}{c}{AEE}&&\multicolumn{3}{c}{Full}\\
		&EST&ASE&$p$-val&&EST&ASE&$p$-val&&EST&ASE&$p$-val\\
		\hline
	AGE & 0.174 & 0.010 & 0.000 && 0.170 & 0.011 & 0.000 && 0.178 & 0.010 & 0.000 \\ 
	WEIGHT & 0.158 & 0.012 & 0.000 && 0.156 & 0.012 & 0.000 && 0.160 & 0.012 & 0.000 \\ 
	HEIGHT & $-$0.043 & 0.012 & 0.000 && $-$0.040 & 0.012 & 0.001 && $-$0.042 & 0.012 & 0.000 \\ 
	SPLIMIT & 0.126 & 0.010 & 0.000 && 0.122 & 0.011 & 0.000 && 0.127 & 0.010 & 0.000 \\ 
	VEHWGT & 0.023 & 0.010 & 0.026 && 0.022 & 0.011 & 0.041 && 0.022 & 0.011 & 0.035 \\ 
	CURBWGT & $-$0.115 & 0.011 & 0.000 && $-$0.115 & 0.012 & 0.000 && $-$0.119 & 0.012 & 0.000 \\ 
	VEHAGE & 0.190 & 0.011 & 0.000 && 0.187 & 0.011 & 0.000 && 0.196 & 0.011 & 0.000 \\ 
	BAGDEPLY & 0.585 & 0.020 & 0.000 && 0.583 & 0.020 & 0.000 && 0.597 & 0.020 & 0.000 \\ 
	PARUSE & $-$1.104 & 0.037 & 0.000 && $-$1.079 & 0.039 & 0.000 && $-$1.134 & 0.038 & 0.000 \\ 
	OCCRACE & $-$0.013 & 0.021 & 0.536 && $-$0.016 & 0.022 & 0.464 && $-$0.016 & 0.021 & 0.465 \\ 
	LANES & $-$0.035 & 0.020 & 0.089 && $-$0.035 & 0.021 & 0.094 && $-$0.036 & 0.021 & 0.083 \\ 
	DRGINV & 0.851 & 0.033 & 0.000 && 0.847 & 0.034 & 0.000 && 0.884 & 0.033 & 0.000 \\ 
	DRIVDIST & $-$0.204 & 0.020 & 0.000 && $-$0.206 & 0.021 & 0.000 && $-$0.213 & 0.021 & 0.000 \\ 
	SURCOND & 0.139 & 0.025 & 0.000 && 0.142 & 0.025 & 0.000 && 0.145 & 0.025 & 0.000 \\ 
	PREVACC & 0.006 & 0.028 & 0.834 && $-$0.003 & 0.029 & 0.928 && $-$0.001 & 0.028 & 0.980 \\ 
	  		\hline\hline
		\end{tabular}
	}
\end{table}

\section{Concluding remarks}
\label{sec:dis}
In this paper, we provide a unified framework of statistical inference based on an extended form of confidence distribution for estimation function, which allows regression analyses of massive data sets, including quantile regression, longitudinal GEE model and Cox regression model. This new paradigm is developed and implemented to achieve parallel and scalable computation via a distributed file system such as Apache Hadoop. Our proposed Rao-CD method serves as the core of methodology, which has been shown to have four salient features: computational scalability, methodological generality, operational robustness and statistical optimality. The most interesting finding is that the proposed Rao-CD method is closely connected to Hansen's GMM and Crowder's optimality. Using this fact, we show that the asymptotic efficiency of our Rao-CD meta estimator is always greater or equal to the estimator obtained by processing the full data once in the context of estimating functions. Furthermore, the implementation of the proposed one-step updating procedure facilitates a fast computation with ignorable numerical approximation errors as the sample size $n \rightarrow \infty$.

Our proposed Rao-CD method has several limitations, including lack of generally adaptive partition rules other than random data partition, and homogeneity assumption. For the former, in this paper, we do not give clear guidelines on how to split the data, which in some practical cases may be important. In general, the strategy of partition is likely to be problem dependent, such as our two data analysis examples. When there is no any prior knowledge about data constructs in hand, it is not yet clear whether the strategy of random data partition would affect power of statistical inference. This is worth further exploration. For the latter, the assumption of homogeneous parameters across sub-datasets may not be fully appropriate in some practical studies. For example, in the FARS data example, as shown by Figure~\ref{fig:cds_month}, the regression coefficients from certain months are different from others. An important future direction of research is to relax the homogeneity assumption to allow heterogeneous parameterization on some part of the model across data batches, while keep other part of the model with common parameters. This relaxation is particularly appealing, when to combine sub-datasets that are collected from different scientific studies and stored in separate data file systems, where integrative data analyses are challenged by a great deal of data heterogeneity. 

To address the situation with heterogeneous regression parameters, the idea in \cite{liu2015multivariate} seems promising, which may provide a way to generalize the Rao-CD meta estimation under heterogeneous inter-dataset parameters. Let $\btheta$ be a $q$-element vector of all distinctive parameters, $q \geq p$. Using known mapping functions $\eta_k: R^{q} \to R^{p}, k = 1, \dots, K$, we are able to yield different versions of $p$-dimensional subvectors with respect to specific sub-datasets. The resulting dimension of the $k$-th estimating function, $\psi_{k\_{sub}}\left(\bW^{(k)}; \eta_k(\btheta)\right)$ remains of $p$-dimension. Moreover, let 
$$\bpsi_{n, \eta}(\bW; \btheta) = \left\{\sqrt{n_1}\psi_{1\_{sub}}\left(\bW^{(1)}; \eta_1(\btheta)\right), \dots, \sqrt{n_K}\psi_{K\_{sub}}\left(\bW^{(K)}; \eta_K(\btheta)\right)  
\right\}^{T},$$
and let $\hat{\mathbb{V}}_{n, \eta} = \text{block-diag}\left\{\hat{\bV}_{n_1, \eta_1}, \dots, \hat{\bV}_{n_K, \eta_K}\right\}$ with $\hat{\bV}_{n_k, \eta_k} = n^{-1}_k\sum_{i=1}^{n_k}\psi\left(\bW_{k, i}; \eta_k(\btheta)\right)\psi^{T}\left(\bW_{k, i}; \eta_k(\btheta)\right)$. Following the Rao-CD meta estimator in~\eqref{gmm:cef}, we have 
$\hat{\btheta}_{rcd} = \arg\min_{\btheta}\left\{\bpsi^{T}_{n, \eta}(\bW; \btheta)\hat{\mathbb{V}}_{n, \eta}^{-1}\bpsi_{n, \eta}(\bW; \btheta) \right\}$.
The meta estimate $\hat{\btheta}_{rcd}$ may be obtained as a solution to the estimating equation 
$\sum_{k=1}^{K}n_k\dot{\eta}^{T}_k(\btheta)\bS^{T}_{n_k}(\eta_k(\btheta))$\\$\times \hat{\bV}^{-1}_{n_k, \eta_k}\psi_{k\_{sub}}\left(\bW^{(k)}; \eta_k(\btheta)\right) = \bm{0}$,
where $\dot{\eta}_k(\btheta) = \partial \eta_k(\btheta)/\partial \btheta$ is the Jacobian matrix for the $k^{th}$ mapping function. The specification of these map functions $\eta_k's$ is problem specific. Also, it is useful to develop some screening procedures to discern possible inter-dataset parameter heterogeneity. One theoretical problem of interest is to investigate semiparametric efficiency for common parameter estimators in the setting of semiparametric models that relaxes homogeneity assumption by allowing heterogeneous nonparametric functions. One of such examples is the Cox proportional hazards model with common regression coefficients but with different baseline hazard functions across sub-datasets.  

This paper focuses on statistical inference on parameters with a fixed dimension. To address the situation where the dimension of parameters increases along with the sample size, it is inevitable to impose regularization to robustify statistical analysis. Developing reliable statistical inferences with certain regularization imposed in meta estimation procedure is of great interest, which has been little investigated in the current literature. In addition, as pointed out in Section~\ref{sec:the} of this paper, the number of computing nodes, K, increases to infinity under a rate constrained with, the sub-dataset size, $m$. To improve this rate, it seems promising to invoke some de-biased methods for estimation on individual sub-datasets. This is also an interesting area of research for the statistical analysis using the MapReduce paradigm. Last but not least, although we have shown some robustness properties for the proposed Rao-CD method, many more remains unknown, such as the issue of breakpoint in terms of data contamination, and the strategy of efficiently realizing low weighting and dilution properties, to improve the robustness of the proposed method so to make it more reliable in practical applications.   

\section*{Acknowledgments}
This research is supported in part by National Institutes of Health grant R01 ES024732 and National Science Foundation grant DMS1513595. 

\bibliographystyle{rss}
\bibliography{ref}

\appendix
\section{Notation and conditions}
Let a neighborhood around $\btheta^*$ be $\mathbb{N}_{\delta}(\btheta^*) = \left\{\btheta: \|\btheta - \btheta^*\|_2 \leq \delta \right\}$. For the $k^{th}$ sub-dataset, denote the expectation of the estimating function $\psi_{k\_{sub}}$ by $\blambda_k(\btheta) = E_{\btheta}\left\{\psi_{k\_{sub}}\left(\bW^{(k)}; \btheta\right) \right\}$, 
the sensitivity matrix by $\bs_k(\btheta) = - \partial \lambda_k(\btheta)/\partial \btheta$ and the scaled distance between estimating functions at two different points by
$$\mathcal{R}_{n_k}(\btheta_1, \btheta_2) =  \frac{\sqrt{n_k}\|
	\psi_{k\_sub}(\bW^{(k)}; \btheta_1) - \psi_{k\_{sub}}(\bW^{(k)}; \btheta_2) - \blambda_k(\btheta_1) + \blambda_k(\btheta_2)\|_2}{1 + \sqrt{n_k}\|\btheta_1 - \btheta_2\|_2}, \hspace{0.2cm} k = 1, \dots, K.$$
The regularity conditions are given as follows:
\begin{itemize}
	\item[(C1)] For $k = 1, \dots, K$, $\blambda_k(\btheta_{k, 0}) = \bm{0}$, where $\btheta_{k, 0} \in \bTheta \subset \mathcal{R}^{p}$ and $\btheta_{k, 0}$ is the true value of the parameter $\btheta$ of interest for the $k^{th}$ sub-dataset.
	
	\item[(C2)] For $k = 1, \dots, K$, assume there exists a positive constant $\delta$, the sensitivity matrix $\bs_{k}(\btheta)$ is first-order continuously differentiable and positive definite for $\btheta \in \mathbb{N}_{\delta}(\btheta_{k, 0})$.
	
	\item[(C3)] Assume an unbiased estimating function, {\em i.e.}, $E\left\{\sum_{k=1}^{K}n_k\bS_{n_k}(\btheta_{k, 0})\bV^{-1}_{n_k}(\btheta_{k, 0})\psi_{k\_{sub}}(\bW^{(k)}; \btheta_{k, 0})\right\} = \bm{0}$, and Godambe information 
	$j_{cd}(\btheta)$ is positive definite for $\btheta \in \cap_{k=1}^{K}\mathbb{N}_{\delta}(\btheta_{k, 0})$.
	
	\item[(C4.0)] For any $0 \leq \varepsilon_{n_k} \to 0$ and $k = 1, \dots, K$, $\sup_{\btheta \in \mathbb{N}_{\varepsilon_{n_k}}(\btheta_{k, 0})}\mathcal{R}_{n_k}(\btheta, \btheta_{k, 0}) = o_p(1)$.
	\item[(C4.1)] For any $0 \leq \varepsilon_{n_k} \to 0$ and $k = 1, \dots, K$, $\sup_{\btheta \in \mathbb{N}_{\varepsilon_{n_k}}(\btheta_{k, 0})}\mathcal{R}_{n_k}(\btheta, \btheta_0) = O_p\left(
	\|\btheta - \btheta_0\|_2\right)$.
	\item[(C4.2)] Assume for any $0 \leq \varepsilon_{n_k} \to 0$, $\sup_{\btheta_1 \in \mathbb{N}_{\varepsilon_{n_k}}(\btheta_{2})}\mathcal{R}_{n_k}(\btheta_1, \btheta_2) = O_p\left(
	\|\btheta_1 - \btheta_2\|_2\right)$, $\btheta_1, \btheta_2 \in \mathbb{N}_{\delta}(\btheta_{k, 0})$.

\end{itemize}
Conditions (C1), (C2) and (C4.0) are regular conditions for general estimating functions that are not necessarily smooth. Specifically, conditions (C1) and (C2) are responsible for the unbiasedness of estimation equations and uniqueness of the estimator obtained from each individual sub-dataset in the Map-step, respectively. These two conditions are commonly assumed in the literature of estimating functions \cite[{\emph{e.g.}}][]{godambe1991estimating, hu2000estimating}. Condition (C4.0) is needed for the asymptotic normality of the estimator in the presence of non-smooth estimating functions; see \cite{pakes1989simulation, newey1994large}, among others. Conditions (C4.1) and (C4.2) are required for the asymptotic distribution of the Rao-CD meta estimator $\hat{\btheta}_{rcd}$. It is easy to see that both conditions (C4.1) and (C4.2) automatically hold if $\psi_{k\_{sub}}(\bW^{(k)}; \btheta)$ is twice continuously differentiable.  Condition (C3) is mild, ensuring the unbiasedness of the aggregated estimating functions. 

\section{Proofs of the main theorems}

\subsection{Proofs of Theorem~\ref{the:cef_con} and part (a) of Theorem~\ref{the:cefk}}
To prove the consistency of the Rao-CD meta estimator $\hat{\btheta}_{rcd}$, we proceed in two steps. The first step is to establish the asymptotic properties of the estimator $\hat{\btheta}_k$ obtained from each individual sub-dataset in the Map-step, and the second step is to show the consistency of the combined estimator $\hat{\btheta}_{rcd}$ in the Reduce-step.

First, we show the consistency and asymptotic normality of each individual estimator $\hat{\btheta}_k$, $k = 1,\cdots, K$. By the law of large number, $\lim_{n_k \to \infty}\psi_{k\_{sub}}\left(\bW^{(k)}; \btheta_k\right) = \blambda_k\left(\btheta_k\right).$
Since $\psi_{k\_{sub}}(\bW^{(k)}; \hat{\btheta}_k) = \bm{0}$ we have $\blambda_k(\hat{\btheta}_k) = \bm{0}$. Using conditions (C1), (C2) and $\btheta_{k, 0} = \btheta_0, k = 1, \dots, K$, we obtain the estimation consistency, $\hat{\btheta}_k \overset{P}{\rightarrow} \btheta_{0}$ as $n_k \to \infty$. 

To show the asymptotic normality, we take the first-order Taylor expansion of the estimating function around $\btheta_0$ 
\begin{eqnarray*}
	\blambda_k(\hat{\btheta}_k) - \blambda_k(\btheta_0) = -\bs_k(\btheta_0)(\hat{\btheta}_k - \btheta_0) + O_p(\|\hat{\btheta}_k - \btheta_0\|_2^2).
\end{eqnarray*}
Applying condition (C4.0), we have 
\begin{eqnarray}
\label{eq: basic}
\|\psi_{k\_{sub}}(\bW^{(k)}; \hat\btheta_k) - \psi_{k\_{sub}}(\bW^{(k)}; \btheta_0)  + \bs_k(\btheta_0)(\hat{\btheta}_k - \btheta_0)\|_2 = o_p\left(n_k^{-1/2} + \|\hat{\btheta}_k - \btheta_0\|_2\right).
\end{eqnarray}

It is known that by the central limit theory,  
\begin{eqnarray*}
\sqrt{n_k}\psi_{k\_{sub}}(\bW^{(k)}; \btheta_{0}) \overset{d}{\rightarrow} \mathcal{N}\left(\bm{0}, \bv_{k}(\btheta_{0})\right), \hspace{0.2cm} \mbox{as} \hspace{0.2cm} n_k\to \infty,
\end{eqnarray*}
where $\bv_{k}(\btheta) = \mbox{Var}\left\{\psi(\bW^{(k)}_{i}; \btheta)\right\}$. Combining the above two equations, 
we obtain 
\begin{eqnarray}
\label{eq:asyk_p}
\sqrt{n_k}\left(\hat{\btheta}_k - \btheta_{0}\right) \overset{d}{\rightarrow} \mathcal{N}\left(\bm{0}, \bj^{-1}_k(\btheta_{0})\right), \hspace{0.2cm}{as} \hspace{0.2cm} n_k\to \infty,
\end{eqnarray} 
where the Godambe information matrix, $\bj_k(\btheta) = \bs^{T}_k(\btheta)\bv^{-1}_k(\btheta)\bs_k(\btheta)$.

Second, we focus on the consistency of $\hat{\btheta}_{rcd}$. Note that $\hat{\btheta}_{rcd}$ is the solution of the following equation,
\begin{eqnarray*}
n^{-1}\sum_{k=1}^{K}n_k\bS^{T}_{n_k}(\hat{\btheta}_{rcd})\hat{\bV}^{-1}_{n_k}\psi_{k\_{sub}}(\bW^{(k)}; \hat{\btheta}_{rcd}) = \bm{0}.
\end{eqnarray*}
Condition (C3) implies that $\hat{\btheta}_{rcd} \overset{p}{\to} \btheta_0 $. 
In fact, letting $C_{n_k}(\btheta; \btheta_k) = \bS^{T}_{n_k}(\btheta)\bV^{-1}_{n_k}(\btheta_k)$, we have
the following orders of the covariances $$\bigg|\mbox{Cov}\left\{C_{n_k}(\btheta; \btheta_k)\psi(\bW^{(k)}_{i}; \btheta), C_{n_k\prime}(\btheta; \btheta_{k^{\prime}})\psi(\bW^{(k^{\prime})}_{ i^{\prime}}; \btheta)\right\}\bigg| 
=\left\{\begin{array}{ll} 
0, & k\neq k^{\prime};\\ 
O(n^{-1}_k), & k= k^{\prime} \hspace{0.2cm}\mbox{and} \hspace{0.2cm} i \neq i^{\prime};\\
O(1), & k= k^{\prime} \hspace{0.2cm}\mbox{and} \hspace{0.2cm} i=i^{\prime}.
\end{array}\right.$$

Employing the law of large number, we have 
\begin{eqnarray*}
\bm{0} &=& n^{-1}\sum_{k=1}^{K}\sum_{i=1}^{n_k}C_{n_k}(\btheta; \btheta_k)\psi(\bW^{(k)}_{i}; \btheta)\mid_{\btheta_k = \hat{\btheta}_k, \btheta = \hat{\btheta}_{rcd}} \nonumber\\
&=& n^{-1}\sum_{k=1}^{K}\sum_{i=1}^{n_k}E\left\{ C_{n_k}(\btheta; \btheta_k)\psi(\bW^{(k)}_{i}; \btheta)\right\}\mid_{\btheta_k = \hat{\btheta}_k, \btheta = \hat{\btheta}_{rcd}} + O_p(n^{-1/2})\nonumber\\
&=& n^{-1}\sum_{k=1}^{K}\sum_{i=1}^{n_k}E\left\{ C_{n_k}(\btheta; \btheta_0)\psi(\bW^{(k)}_{i}; \btheta)\right\}\mid_{\btheta = \hat{\btheta}_{rcd}} + O_p\left[n^{-1}\left(\sum_{k=1}^Kn^{1/2}_k\right)\right],
\end{eqnarray*}
where the third equation holds due to the asymptotic normality given in~(\ref{eq:asyk_p}).
Combining the above equation with condition (C3), we establish the estimation consistency, $\hat{\btheta}_{rcd} \overset{p}{\rightarrow}\btheta_0$ as $m \to \infty$.
\blackbox

\subsection{Proofs of Theorem~\ref{the:cef_norm} and part (b) of Theorem~\ref{the:cefk}}
Now, we show the asymptotic normality of $\hat{\btheta}_{rcd}$. Note that 
\begin{eqnarray}
\label{eq:asycee1}
&&n^{-1}\sum_{k=1}^{K}n_kC_{n_k}(\hat{\btheta}_{rcd}; \hat{\btheta}_k)\psi_{k\_sub}(\bW^{(k)}; \hat{\btheta}_{rcd}) - n^{-1}\sum_{k=1}^{K}n_kC_{n_k}(\btheta_0; \hat{\btheta}_k)\psi_{k\_{sub}}(\bW^{(k)}; \btheta_{0})\nonumber\\
&&\hspace{0.5cm} = n^{-1}\sum_{k=1}^{K}n_kC_{n_k}(\hat{\btheta}_{rcd}; \hat{\btheta}_k)
\left\{\psi_{k\_{sub}}(\bW^{(k)}; \hat{\btheta}_{rcd}) - 
\psi_{k\_{sub}}(\bW^{(k)}; \btheta_0)\right\}\nonumber\\
&&\hspace{3.5cm} + n^{-1}\sum_{k=1}^{K}n_k\left\{\bS_{n_k}(\hat{\btheta}_{rcd}) - \bS_{n_k}(\btheta_{0})\right\}^{T}\hat{\bV}^{-1}_{n_k}
\psi_{k\_{sub}}(\bW^{(k)}; \btheta_{0})\nonumber\\
&&\hspace{0.5cm} = -n^{-1}\sum_{k=1}^{K}n_kC_{n_k}(\hat{\btheta}_{rcd}; \hat{\btheta}_k)\bs_{k}(\btheta_0)(\hat{\btheta}_{rcd} - \btheta_0) + O_p\left(n^{-1}(\sum_{k=1}^{K}n_k^{1/2})\|\hat{\btheta}_{rcd} - \btheta_0\|_2\right)\nonumber\\
&&\hspace{0.5cm} = -n^{-1}\sum_{k=1}^{K}n_kJ_{n_k}(\btheta_0)(\hat{\btheta}_{rcd} - \btheta_0) + O_p\left(n^{-1}(\sum_{k=1}^{K}n_k^{1/2})\|\hat{\btheta}_{rcd} - \btheta_0\|_2 + \|\hat{\btheta}_{rcd} -\btheta_0\|^2_2\right),
\end{eqnarray}
where the second equality follows from condition (C4.1).
On the other hand,  
\begin{eqnarray}
\label{eq:cov}
&&\mbox{Cov}\left\{C_{n_k}(\btheta_0; \btheta_0)\psi(\bW^{(k)}_{i};\btheta_0), C_{n_k}(\btheta_0; \btheta_0)\psi(\bW^{(k)}_{i'};\btheta_0)\right\}\nonumber\\
&&\hspace{-0.5cm} = 
E\left[ \left\{ n^{-1}_k\sum_{i=1}^{n_k}\bS_{k, i}(\btheta_0) \right\}^{T}
\left\{ n^{-1}_k\sum_{i=1}^{n_k}\psi(\bW^{(k)}_{i}; \btheta_0)\psi^{T}(\bW^{(k)}_{i}; \btheta_0) \right\}^{-1}\right. \nonumber\\
&&\hspace{0cm}\left. \times \psi(\bW^{(k)}_{i}; \btheta_0)\psi^{T}(\bW^{(k)}_{i'}; \btheta_0) 
\left\{ n^{-1}_k\sum_{i=1}^{n_k}\psi(\bW^{(k)}_{i}; \btheta_0)\psi^{T}(\bW^{(k)}_{i}; \btheta_0) \right\}^{-1}\left\{ n^{-1}_k\sum_{i=1}^{n_k}\bS_{k, i}(\btheta_0) \right\}\right]\nonumber\\
&&\hspace{-0.5cm} =\left\{\begin{array}{ll} \left\{\bs^{T}_{k}(\btheta_0)\bv_{k}^{-1}(\btheta_0)\bs_{k}(\btheta_0)
\right\}\left(1 + O(n_k^{-1/2})\right), & i = i',\\
O(n^{-2}_k), & i\neq i',
\end{array}
\right.
\end{eqnarray}
where $\bS_{k, i}(\btheta)$ is the sample version of $\bs_{k}(\btheta)$, {\em i.e.} $\bS_{n_k}(\btheta) = n^{-1}_k\sum_{i=1}^{n_k}\bS_{k, i}(\btheta)$.

By the results in~(\ref{eq:cov}), we have 
\begin{eqnarray}
\label{eq:asycee2}
&&\hspace{0.2cm}n^{-1/2}\sum_{k=1}^{K}n_kC_{n_k}(\btheta_0; \hat{\btheta}_k)\psi_{k\_{sub}}(\bW^{(k)}; \btheta_{0})\nonumber\\ 
&&= n^{-1/2}\sum_{k=1}^{K}n_kC_{n_k}(\btheta_0; \btheta_0)\psi_{k\_{sub}}(\bW^{(k)}; \btheta_{0}) + O_p(n^{-1/2}K) \nonumber\\
&&= n^{-1/2}\sum_{k=1}^{K}\sum_{i=1}^{n_k}\bs^{T}_{k}(\btheta_{0})\bv^{-1}_{k}(\btheta_0)\psi(\bW^{(k)}_i; \btheta_0) + o_p(1),
\end{eqnarray} 
where the last equation holds under the condition of $K = O(n^{1/2 - \delta} )$.

Combining equations (\ref{eq:asycee1})--(\ref{eq:asycee2}) and applying the central limit theory, we obtain the asymptotic normality of $\hat{\btheta}_{rcd}$,
\begin{eqnarray*}
\sqrt{n}\left(\hat{\btheta}_{rcd} - \btheta_0\right)  \stackrel{d}{\rightarrow} \mathcal{N}\left(\bm{0}, \bj_{cd}^{-1}(\btheta_0)\right), \hspace{0.5cm}\text{as} \hspace{0.2cm} m = \min_{k}n_k \to \infty,
\end{eqnarray*}
where $\bj_{cd}(\btheta) = \lim_{m\to \infty}n^{-1}\sum_{k=1}^{K} n_k\bj_k(\btheta)$, with $\bj_k(\btheta) = \bs^{T}_{k}(\btheta)\bv_{k}^{-1}(\btheta)\bs_{k}(\btheta)$.
\blackbox

\subsection{Proof of Proposition~\ref{prop:crowder}}
Under conditions (C1) and (C2), it follows from the central limit theory that $\psi_{k\_{sub}}(\bW^{(k)}; \btheta_0) = O_p(n_k^{-1/2})$. According to the definition of $\Psi_{R}(\btheta)$ in~(\ref{eq:cef}), we have
\begin{eqnarray*}
\Psi_{R}(\btheta_0) - \Psi^*_c(\btheta_0) & = & n^{-1/2}\sum_{k=1}^Kn_k\left\{C_{n_k}(\btheta_0; \hat{\btheta}_k) - C_k(\btheta_0) \right\}\psi_{k\_sub}(\bW^{(k)}; \btheta_0)\\
& = & n^{-1/2}\sum_{k=1}^Kn_k\left\{\bS_{n_k}(\btheta_0) - \bs_k(\btheta_0) \right\}^{T}\hat{\bV}^{-1}_{n_k}\psi_{k\_sub}(\bW^{(k)}; \btheta_0)\\
&&\hspace{0.6cm} + n^{-1/2}\sum_{k=1}^Kn_k\bs^{T}_k(\btheta_0)\left\{\hat{\bV}^{-1}_{n_k} - \bv^{-1}_k(\btheta_0) \right\}\psi_{k\_sub}(\bW^{(k)}; \btheta_0)\\
& = & O_p(n^{-1/2}K),
\end{eqnarray*}
where the third equation holds from the law of large number and the asymptotic formula in~(\ref{eq:asyk_p}). Moreover, by the condition that $K = O(n^{1/2 - \delta})$ with $\delta \in (0, 0.5]$, Proposition~\ref{prop:crowder} follows.
\blackbox

\subsection{Proof of Theorem~\ref{the:cd&cef}}
Let $\hat{\btheta}_{wcd}$ be a solution of the following estimating equation,
\begin{eqnarray}
\label{eq:wcdest}
n^{-1}\sum_{k=1}^{K}n_kC_{n_k}(\hat{\btheta}_k; \hat{\btheta}_k)\hat{\bS}_{n_k} \left(\hat{\btheta}_{k} - \hat{\btheta}_{wcd}\right) = \bm{0}.
\end{eqnarray}
We begin the proof with a lemma.
\begin{lemma}
	\label{lem:cd}
	If all conditions of Theorem \textbf{3} hold, we have (i) $\hat{\btheta}_{wcd} \overset{p}{\rightarrow} \btheta_0$, and 
	(ii) $\sqrt{n}\left(\hat{\btheta}_{wcd} - \btheta_0\right) \stackrel{d}{\rightarrow} \mathcal{N}\left(\bm{0}, \bj^{-1}_{cd}(\btheta_0)\right)$, as $m = \min_{k}{n_k} \to \infty$. 
\end{lemma}
\begin{proof}
According to equation~(\ref{eq:wcdest}), we rewrite~(\ref{eq:wcdest}) as follows: 
\begin{eqnarray*}
	\bm{0} = n^{-1}\sum_{k=1}^{K}n_kC_{n_k}(\hat{\btheta}_k; \hat{\btheta}_k)\hat{\bS}_{n_k} \left(\hat{\btheta}_{k} - \btheta_0\right)  + n^{-1}\sum_{k=1}^{K}n_kC_{n_k}(\hat{\btheta}_k; \hat{\btheta}_k)\hat{\bS}_{n_k}\left(\btheta_0 - \hat{\btheta}_{wcd}\right).
\end{eqnarray*}
It follows that 
\begin{eqnarray*}
n^{1/2}\left(\hat{\btheta}_{wcd} - \btheta_0\right)&=& \left\{n^{-1}\sum_{k=1}^{K}n_k\bJ_{n_k}(\hat{\btheta}_k) \right\}^{-1}\left\{n^{-1/2}\sum_{k=1}^{K}n_kC_{n_k}(\hat{\btheta}_k; \hat{\btheta}_k)\hat{\bS}_{n_k} \left(\hat{\btheta}_{k} - \btheta_0\right) \right\}\\
&=&\left\{n^{-1}\sum_{k=1}^{K}n_k\bJ_{n_k}(\hat{\btheta}_k) \right\}^{-1}\left[n^{-1/2}\sum_{k=1}^{K}n_k\bs^{T}_{k}\bv^{-1}_{k}\left\{\bs_{k}(\btheta_0) \left(\hat{\btheta}_{k} - \btheta_0\right)\right\} \right] + O_p(Kn^{-1/2})\\
&=&\left\{n^{-1}\sum_{k=1}^{K}n_k\bJ_{n_k}(\hat{\btheta}_k) \right\}^{-1}\left\{n^{-1/2}\sum_{k=1}^{K}\sum_{i=1}^{n_k}\bs^{T}_{k}\bv^{-1}_{k}\psi(\bW^{(k)}_{i}; \btheta_0) \right\} + O_p(Kn^{-1/2}),
\end{eqnarray*}
where $\bJ_{n_k}(\btheta) = \bS^{T}_{n_k}(\btheta)\bV^{-1}_{n_k}(\btheta)\bS_{n_k}(\btheta)$ and the third equation holds under condition (C4.1). Given the condition of $K = O(n^{1/2-\delta})$, $\delta \in (0, 1/2]$, part (i) holds by the law of large number, and part (ii) follows from 
the central limit theorem.
\end{proof}
\blackbox

Now we turn to prove Theorem~\ref{the:cd&cef}.
According to equation (\ref{eq:cef}), we have
\begin{eqnarray*}
\bm{0} &=& n^{-1}\sum_{k=1}^{K}n_kC_{n_k}(\hat{\btheta}_k; \hat{\btheta}_k)\psi_{k\_{sub}}(\bW^{(k)}; \hat{\btheta}_{rcd}) + n^{-1}\sum_{k=1}^{K}n_k\left\{\bS_{n_k}(\hat{\btheta}_{rcd}) - \hat{\bS}_{n_k}\right\}^{T}\hat{\bV}^{-1}_{n_k}\psi_{k\_{sub}}(\bW^{(k)}; \hat{\btheta}_{rcd}) \\
&=& n^{-1}\sum_{k=1}^{K}n_kC_{n_k}(\hat{\btheta}_k; \hat{\btheta}_k)\left\{\psi_{k\_{sub}}(\bW^{(k)}; \hat{\btheta}_{wcd}) - \bs_k(\hat{\btheta}_{wcd})\left(\hat{\btheta}_{rcd} - \hat{\btheta}_{wcd}\right)\right\}\\ 
&&\hspace{3cm} + O_p\left\{\|\hat{\btheta}_{rcd} - \hat{\btheta}_{wcd}\|_2^2 + n^{-1}(\sum_{k=1}^{K}n_k^{1/2}) \|\hat{\btheta}_{rcd} - \hat{\btheta}_{wcd}\| + Kn^{-1}\right\},
\end{eqnarray*}
where the second equation follows from condition (C4.2). Then it is sufficient to show that the following term is asymptotically negligible. That is,
\begin{eqnarray*}
&&n^{-1}\sum_{k=1}^{K}n_kC_{n_k}(\hat{\btheta}_k; \hat{\btheta}_k)\bs_k(\hat{\btheta}_{wcd})\left(\hat{\btheta}_{rcd} - \hat{\btheta}_{wcd}\right)\\
&&\hspace{-0.5cm} = n^{-1}\sum_{k=1}^{K}n_kC_{n_k}(\hat{\btheta}_k; \hat{\btheta}_k)\psi_{k\_{sub}}(\bW^{(k)}; \hat{\btheta}_{wcd}) + O_p\left(n^{-1}(\sum_{k=1}^{K}n_k^{1/2}) \|\hat{\btheta}_{rcd} - \hat{\btheta}_{wcd}\| + Kn^{-1}\right)\\
&&\hspace{-.5cm} = -n^{-1}\sum_{k=1}^{K}n_kC_{n_k}(\hat{\btheta}_k; \hat{\btheta}_k)\bs_k(\hat{\btheta}_{k})\left(\hat{\btheta}_k -  \hat{\btheta}_{wcd}\right)
+ O_p\left(n^{-1}(\sum_{k=1}^{K}n_k^{1/2}) \|\hat{\btheta}_{rcd} - \hat{\btheta}_{wcd}\| + Kn^{-1}\right)\\
&&\hspace{-.5cm} = n^{-1}\sum_{k=1}^{K}n_kC_{n_k}(\hat{\btheta}_k; \hat{\btheta}_k)\left\{\hat{\bS}_{n_k} - \bs_k(\hat{\btheta}_{k})\right\}\left(\hat{\btheta}_k -  \hat{\btheta}_{wcd}\right)
+ O_p\left(n^{-1}(\sum_{k=1}^{K}n_k^{1/2}) \|\hat{\btheta}_{rcd} - \hat{\btheta}_{wcd}\| + Kn^{-1}\right)\\
&&\hspace{-0.5cm} = O_p(Kn^{-1}),
\end{eqnarray*}
where the last equation holds from Lemma~\ref{lem:cd}.
\blackbox

\subsection{Proof of Theorem~\ref{the:var} }
By some simple algebra, we have
	\begin{eqnarray}
	\label{eq:l1}
	&&\sum_{k=1}^{K}n_k\left\{\bJ_{n_k}(\hat{\btheta}_k) - \bj_k(\btheta_0)
	\right\}\nonumber\\
	&& = \sum_{k=1}^{K}n_k\left\{\bS_{n_k}(\hat{\btheta}_k)
	- \bS_{n_k}({\btheta}_0)\right\}^{T}C^{T}_{n_k}(\hat{\btheta}_k; \hat{\btheta}_k)
	 + \sum_{k=1}^{K}n_k\left\{\bS_{n_k}({\btheta}_0)
	- \bs_{k}({\btheta}_0)\right\}^{T}C^{T}_{n_k}(\hat{\btheta}_k; \hat{\btheta}_k)\nonumber\\
	&& + \sum_{k=1}^{K}n_k\bs^{T}_{k}({\btheta}_0)\left\{\bV^{-1}_{n_k}(\hat{\btheta}_k) - \bv^{-1}_{k}(\btheta_0)\right\}\bS_{n_k}(\hat{\btheta}_k) + \sum_{k=1}^{K}n_k\bs^{T}_{k}({\btheta}_0)\bv^{-1}_{k}(\btheta_0)\left\{
	\bS_{n_k}(\hat{\btheta}_k) - \bs_{k}({\btheta}_0)\right\}\nonumber\\ 
	&&\overset{def}{=} I_1 + I_2 + I_3 + I_4.
	\end{eqnarray}
	Using expression (\ref{eq: basic}) and condition (C4.1), we obtain  
	\begin{eqnarray*}
		\bS_{n_k}(\hat{\btheta}_k) &=& 
		\bS_{n_k}(\btheta_0) - 
		n^{-2}_k\sum_{i, j=1}^{n_k}\bm{M}_{ij}(\btheta_0) + O_p(n_k^{-1}),\\
		\bV_{n_k}(\hat{\btheta}_k) &=& n^{-1}_k\sum_{i=1}^{n_k}\left\{
		\psi(\bW^{(k)}_{i}; \btheta_0) - \bs^{T}_{k}(\btheta_0)(\hat{\btheta}_k - \btheta_0) +O_p(\|\hat{\btheta}_k - \btheta_0\|_2) 
		\right\}^{\otimes 2}\\
		&=&\bV_{n_k}(\btheta_0) + O_p(n_k^{-1}),
	\end{eqnarray*}
	where $\bm{M}_{ij}(\btheta_0)$ is a $p \times p$ dimensional matrix, whose $(l,m)$-th element is 
	$\frac{d \{\bS_{k, i}(\btheta)\}_{lm}}{d \btheta^{T}}\mid_{\btheta = \btheta_0}\bs^{-1}_{k}(\btheta_0)\psi(\bW^{(k)}_{j}; \btheta_0)$, with $\{\bS_{k, i}(\btheta)\}_{lm}$ being the $(l, m)$-th component of $\bS_{k, i}(\btheta)$, $\bS_{n_k}(\btheta) = n^{-1}_k\sum_{i=1}^{n_k}\bS_{k, i}(\btheta)$.
	
	It follows that
	\begin{eqnarray}
	\label{eq:ls}
	n^{-1}I_1 &=& -n^{-1}\sum_{k=1}^{K}n_k\left\{n^{-2}_k\sum_{i, j=1}^{n_k}\bm{M}_{ij}(\btheta_0) + O_p(n_k^{-1})
	\right\}\nonumber\\
	&&\hspace{1.8cm} \times \left\{\hat{\bV}^{-1}_{n_k} - \bv^{-1}_{k}(\btheta_0) + \bv^{-1}_{k}(\btheta_0)
	\right\} \left\{\hat{\bS}_{n_k} - \bs_{k}(\btheta_0) + \bs_{k}(\btheta_0)\right\}\nonumber\\
	& = &O_p(n^{-1/2} + Kn^{-1}); \nonumber\\
	n^{-1}I_2 &=&n^{-1}\sum_{k=1}^{K}n_k\left\{n^{-1}_k\sum_{i, =1}^{n_k}\bS_{k, i}(\btheta_0)- \bs_{k}(\btheta_0)
	\right\}^{T}\bv^{-1}_{k}(\btheta_0)\bs_{k}(\btheta_0) (1 + o_p(1))\nonumber\\
	&=& n^{-1}\sum_{k=1}^{K}\sum_{i=1}^{n_k}\left\{\bS^{T}_{k, i}(\btheta_0)\bv^{-1}_{k}(\btheta_0)\bs_{k}(\btheta_0) - \bj_k(\btheta_0)
	\right\}(1  + o_p(1)) = O_p(n^{-1/2}).
	\end{eqnarray}
	Similarly,
	\begin{eqnarray}
	\label{eq:lv}
	n^{-1}I_3 & = &-n^{-1}\sum_{k=1}^{K}n_k\bs^{T}_{k}(\btheta_0)\hat{\bV}^{-1}_{n_k}
	\left\{\hat{\bV}_{n_k} - \bv_{k}(\btheta_0)\right\}
	\bv^{-1}_{k}(\btheta_0)\bS_{n_k}(\btheta_0)\nonumber\\
	&=&-n^{-1}\sum_{k=1}^{K}n_k\bs^{T}_{k}(\btheta_0)\bv^{-1}_{k}({\btheta}_0)
	\left\{n^{-1}_k\sum_{i=1}^{n_k}\psi(\bW^{(k)}_{i}; \btheta_0)\psi^{T}(\bW^{(k)}_{i}; \btheta_0) - \bv_{k}(\btheta_0)\right\}\nonumber\\
	&&\hspace{2cm}\times \bv^{-1}_{k}(\btheta_0)\bs_{k}(\btheta_0) + O_p(Kn^{-1})\nonumber\\
	& = &O_p(n^{-1/2} + Kn^{-1});\nonumber\\
	n^{-1}I_4 &=& O_p(n^{-1/2} + Kn^{-1}).
	\end{eqnarray}
	Combining above results in (\ref{eq:l1})--(\ref{eq:lv}), Theorem~\ref{the:var} follows.
\blackbox 

\subsection{Proof of Theorem~\ref{the:corie}}
Note that $\psi_{k\_{sub}}(\bW^{(k)}; \btheta_0)$, $k=1,\cdots, K$ are independent. Denote $p_{n_k} = n_k/n$, and 
two $p\times pK$ dimensional matrices $\bH_{n1}$ and $\bH_{n2}$ by $$\bH_{n1} = (\sqrt{p_{n_1}}I_p, \dots, \sqrt{p_{n_K}}I_p)^{T},
\hspace{0.2cm} \bH_{n2} = \left(\sqrt{p_{n_1}}\bs^{T}_{1}(\btheta_0)\bv^{-1}_{1}(\btheta_0), \dots, \sqrt{p_{n_K}}\bs^{T}_{K}(\btheta_0)\bv^{-1}_{K}(\btheta_0)\right)^{T},$$
where $I_p$ is a $p\times p$ identity matrix.
Then, as $m=\min_k\{n_k\} \rightarrow \infty$,
\begin{eqnarray*}
\bH_{n1}^{T}\bpsi_{n}(\bW; \btheta_0) &\overset{d}{\rightarrow}& \mathcal{N}\left(\bm{0}, \bv(\btheta_0)\right),\nonumber\\
\bH_{n2}^{T}\bpsi_{n}(\bW; \btheta_0) &\overset{d}{\rightarrow}& \mathcal{N}\left(\bm{0}, \bj_{cd}(\btheta_0)\right),
\end{eqnarray*}
where $\bv(\btheta) = \lim_{m \to \infty}n^{-1}\sum_{k=1}^{K}n_k\bv_{k}(\btheta)$ and $\bj_{cd}(\btheta) = \lim_{m\to \infty}n^{-1}\sum_{k=1}^{K}n_k
\bj_{k}(\btheta_0)$. 
Also
$$\mbox{Cov}\left\{\bH_{n1}^{T}\bpsi_{n}(\bW; \btheta_0), \bH_{n2}^{T}\bpsi_n(\bW; \btheta_0)\right\} \to\bs(\btheta_0), \hspace{0.2cm} \text{as} \hspace{0.2cm} m \to \infty, $$
where $\bs(\btheta) = \lim_{m\to \infty}n^{-1}\sum_{k=1}^{K}n_k\bs_{k}(\btheta)$. 
It follows that jointly
\begin{eqnarray*}
&&\begin{pmatrix}
\bH_{n1}^{T}\bpsi_{n}(\bW; \btheta_0)\\
\bH_{n2}^{T}\bpsi_{n}(\bW; \btheta_0)\\
\end{pmatrix} \stackrel{d}{\rightarrow} 
\mathcal{N}\left(\bm{0}, \begin{pmatrix}
\bv(\btheta_0)&\bs(\btheta_0)\\
\bs(\btheta_0)&\bj_{cd}(\btheta_0)\\
\end{pmatrix}\right), \hspace{0.2cm} \text{as} \hspace{0.2cm} m \to \infty.
\end{eqnarray*}
It is easy to derive the conditional variance of the following form:
$$\mbox{Var}\left\{\bH_{n2}^{T}\bpsi_{n}(\bW; \btheta_0) \mid \bH_{n1}^{T}\bpsi_{n}(\bW; \btheta_0) = \bm{0}\right\} = \bj_{cd}(\btheta_0) - \bj(\btheta_0).
$$
Thus, Theorem~\ref{the:corie} follows.
\blackbox

\end{document}